\documentclass{aa}
\usepackage[T1]{fontenc}
\usepackage{graphicx}
\usepackage{txfonts}
\usepackage{multirow}
\usepackage{subfig}
\usepackage{epstopdf}
\usepackage{url}
\usepackage{amsmath}
\usepackage{hyperref}
\usepackage{cleveref}
\usepackage{soul}
\usepackage{verbatim}
\usepackage[switch]{lineno}
\hypersetup{
    colorlinks=true,
    linkcolor=blue,
    filecolor=magenta,      
    urlcolor=cyan,
    citecolor=blue,
}
\begin{document}

\title{Multiple accelerated particle populations in the Cygnus Loop with \textit{Fermi}-LAT}

\author{A. Tutone
\inst{1}\fnmsep\inst{2}
\and
J. Ballet
\inst{3}
\and
F. Acero
\inst{3}
\and
A. D'Aì\inst{2}
\and
G. Cusumano\inst{2}
}

\institute{Università degli studi di Palermo, Dipartimento di Fisica e Chimica, Via Archirafi 36 - 90123 Palermo, Italy
\and
INAF/IASF Palermo, Via Ugo La Malfa 153, I-90146 Palermo, Italy
\and
AIM, CEA, CNRS, Université Paris-Saclay, Université de Paris, F-91191 Gif sur Yvette, France
}

\date{Received ; accepted }

\abstract
{The Cygnus Loop (G74.0-8.5) is a very well-known nearby supernova remnant (SNR) in our Galaxy. Thanks to its large size, brightness, and angular offset from the Galactic plane, it has been studied in detail from radio to $\gamma$-ray emission. The $\gamma$ -rays probe the populations of energetic particles and their acceleration mechanisms at low shock speeds.}
{We present an analysis of the $\gamma$-ray emission detected by the Large Area Telescope on board the \textit{Fermi Gamma-ray Space Telescope} over 11 years in the region of the Cygnus Loop.}
{We performed detailed morphological and spectral studies of the $\gamma$-ray emission toward the remnant from 100 MeV to 100 GeV  and compared it with X-ray, UV, optical, and radio images. The higher statistics with respect to the previous studies enabled us to decompose the emission from the remnant into two morphological components to model its nonthermal multiwavelength emission.}
{The extended $\gamma$-ray emission is well correlated with the thermal X-ray and UV emission of the SNR. Our morphological analysis reveals that a model considering two contributions from the X-ray and the UV emission regions is the best description of the $\gamma$-ray data. Both components show a curved spectrum, but the X-ray component is softer and more curved than the UV component, suggesting a different physical origin. The multiwavelength modeling of emission toward the SNR suggests that the nonthermal radio and $\gamma$-ray emission associated with the UV component is mostly due to the reacceleration of preexisting cosmic rays by radiative shocks in the adjacent clouds, while the nonthermal emission associated with the X-ray component arises from freshly accelerated cosmic rays.}
{}

   \keywords{ acceleration of particles – shock waves – ISM: cosmic rays – ISM: supernova remnants }

\maketitle

\section{Introduction}
It is widely accepted that supernova remnants (SNRs) accelerate cosmic rays (CRs) through their fast shock waves that propagate into the interstellar medium (ISM). In particular, SNRs are characterized by the diffusive shock acceleration (DSA) process~\citep{Bell_1978a, Bell_1978b, Blandford_1978, Malkov_2001} that results in nonthermal emission observed from radio to $\gamma$-rays. Strong $\gamma$-ray emission has been observed by the \textit{Fermi} Large Area Telescope (LAT) and the \textit{AGILE} satellite in SNRs interacting with interstellar material. These SNRs are typically evolved and extended intermediate-age ($>$ 10 kyr) remnants interacting with molecular clouds, with a characteristic high-energy break between 1 and 20 GeV~\citep{Giuliani_2011, Ackermann_2013}. The spectrum of these sources can be explained by $\pi^0$ decay emission of accelerated CRs protons in the shocks of SNRs, or alternatively, by the reacceleration of ambient Galactic CRs inside the shock-compressed clouds~\citep{Uchiyama_2010}. The study of intermediate-age SNRs is therefore crucial for understanding the CR acceleration at modest shock speeds (at which the bulk of GeV CRs are accelerated) and the importance of the reacceleration mechanism.

A prototypical intermediat-age SNR is the Cygnus Loop. It is about $21$ kyr old at a distance of $735$ pc that was derived from Gaia parallax measurements of several stars~\citep{Fesen_2018}. It is slightly aspherical, with minor and major axes of $37$ and $47$ pc, E-W and N-S, respectively. Its large size ($\sim 3^{\circ}$) and angular offset from the Galactic plane ($b \sim -8.5^{\circ}$) ensured that this remnant has been widely studied from radio~\citep{Uyaniker_2004, Sun_2006, Loru_2021}, infrared~\citep{Sankrit_2014, Koo_2016}, optical~\citep{Katsuda_2016, Fesen_2018}, UV~\citep{Blair_2002, Kim_2014}, X-ray~\citep{Katsuda_2011, Oakley_2013}, and $\gamma$-rays~\citep{Katagiri_2011, Acero_2016}. The SNR has an approximate shell morphology, with a prominent limb in the northeast region, a blow-out in the south, and several filaments in the north-central region. Several studies~\citep{Levenson_1998, Uchida_2009} and hydrodynamical simulations~\citep{Fang_2017} have shown that the Cygnus Loop properties and morphology are consistent with a scenario of a supernova (SN) explosion taking place in a wind-blown cavity created by the progenitor star. However, very recently,~\citet{Fesen_2018} have proposed that the Cygnus Loop evolved in a low-density region with discrete interstellar clouds in its vicinity: a dense molecular cloud to its west and northwest, and smaller clouds in the east and northeast regions.

A previous analysis of the Cygnus Loop region in the $\gamma$-ray band was performed by~\citet{Katagiri_2011}, who modeled it with a ring with inner and outer radii of $0.7^{\circ} \pm 0.1^{\circ}$ and $1.6^{\circ} \pm 0.1^{\circ}$. They described its emission with a log-normal (LP for LogParabola) spectrum,
\begin{equation}
    \frac{dN}{dE} = N_0 \; \left(\frac{E}{E_b}\right)^{-(\alpha+\beta\log(E/E_b))}
\label{eq:logparabola}
.\end{equation}
In this work we analyze $\sim 11$ years of \textit{Fermi}-LAT data.
This represents an improvement of a factor 5 with respect to the previous study by~\citet{Katagiri_2011}, providing us with unprecedented sensitivity to study both spatial and spectral features of the $\gamma$-ray emission from the Cygnus Loop.

In Section~\ref{section:Obs} we briefly describe the observations and data reduction. Our morphological and spectral analysis is reported in Section~\ref{section:Analysis}. The origin of the $\gamma$-ray emission is discussed in Section~\ref{section:Results}. Finally, the conclusions are summarized in Section~\ref{section:Conclusions}. 

\section{Observations and data reduction}
\label{section:Obs}

Our primary goal is to model the $\gamma$-ray emission from the Cygnus Loop. To this end, we correlated the $\gamma$-ray data with templates from other wavelengths that are characteristic of distinct physical processes.

\begin{figure*}
\centering
\includegraphics[width=0.9\textwidth]{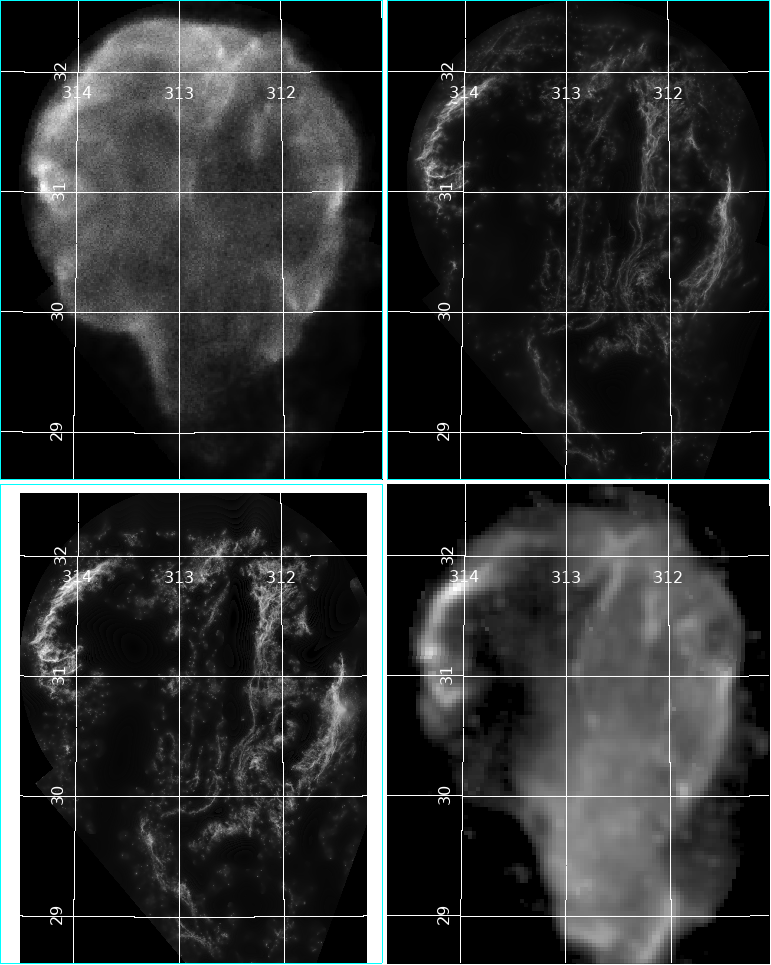}
\caption{Templates used to fit the $\gamma$-ray data in celestial coordinates. Top left: X-rays (ROSAT, Sect.~\ref{obs:Xray}). Top right: UV (GALEX NUV, Sect.~\ref{obs:UV}). Bottom left: Optical (DSS2 red, Sect.~\ref{obs:Opt}). Bottom right: Radio (11 cm, Effelsberg, Sect.~\ref{obs:Radio}). All maps are aligned and are shown in square root scaling from 0 to the maximum.}
\label{fig:multilambda}
\end{figure*}

\subsection{Gamma-ray band}
\label{obs:Gamma}

The LAT is the main instrument on board the \textit{Fermi} satellite. It is a pair-conversion instrument sensitive to $\gamma$-rays in the energy range from $30$ MeV to more than $1$ TeV~\citep{Atwood_2009}.
For this analysis, we used more than 11 years (from August 4, 2008, to October 28, 2019) of \textit{Fermi}-LAT P8R3 data \citep{LAT13_P8, LAT18_P305}. The region of interest (ROI) is $10^\circ \times 10^\circ$ and aligned with Galactic coordinates that are centered on the Cygnus Loop (R.A.=20h50m51, Dec=30$^\circ$34$'$06$\arcsec$, equinox J2000.0). The relatively small ROI size was chosen to avoid the strong diffuse emission from the Galactic plane itself.
To analyze the $\gamma$-ray data, we used version $1.2.1$ of the \texttt{Fermitools} and version $0.18$ of the \texttt{Fermipy} package, which are publicly available from the Fermi Science Support Center (FSSC)\footnote{The Science Tools package and supporting documents are distributed by the Fermi Science Support Center and can be accessed at \href{http://fermi.gsfc.nasa.gov/ssc/data/analysis/software/}{http://fermi.gsfc.nasa.gov/ssc/data/analysis/software/}}.

We selected $\gamma$-rays in the $0.1 - 100$ GeV energy range. Because our analysis relies on morphology, we selected events with good point-spread function (PSF), that is, with{\it } good angular resolution: below 316 MeV, we selected data with \texttt{Event Type} PSF2 and PSF3; between 316 MeV and 1 GeV, we added PSF1 events; above 1 GeV, we used all events including the PSF0 \texttt{Event Type}. The bright $\gamma$-ray emission from the Earth's atmosphere was greatly reduced by selecting events within 90$\degr$ from the local zenith below 316 MeV and within 105$\degr$ of the zenith above 316 MeV. We also applied a good time interval (GTIs) selection on the data using the quality flag \texttt{DATA\_QUAL > 1} and the instrument in science configuration (\texttt{LAT\_CONFIG == 1}). We used the \texttt{CLEAN} event class selection and version \texttt{P8R3\_CLEAN\_V2} of the instrument response functions.

To describe the $\gamma$-ray emission around the Cygnus Loop, we performed a binned likelihood analysis for which the pixel size was set to $0.05^{\circ}$. We also used ten energy bins per decade and summed the log-likelihood over the \texttt{Event Type} selections. We included in the model all the background sources from the 4FGL catalog \citep{Abdollahi_2020} within $13^{\circ}$ of the ROI center. We used the \texttt{gll\_iem\_v07.fits} model to describe the Galactic diffuse emission and the tabulated model \texttt{iso\_P8R3\_CLEAN} to describe the isotropic emission, using the appropriate template for each \texttt{Event Type} selection. We included the effects of energy dispersion on all model components. The only exception was the isotropic emission, which was obtained in data space.

\subsection{X-ray band}
\label{obs:Xray}

The X-ray emission is a good tracer of the shocked gas at densities $< 1$ cm$^{-3}$ and temperatures of a few $10^6$ K occupying most of the SNR interior. Because the Cygnus Loop is very large (hard to mosaic with the current generation of X-ray instruments) and its emission is very soft (peaking below the C edge at 284 eV), the image from the \textit{ROSAT} survey \citep{Aschenbach_1999} remains the best reference.
We obtained the full band image (0.1 -- 2.4 keV) from SkyView\footnote{\url{https://skyview.gsfc.nasa.gov}}. We removed by eye disks of 0.1$\degr$ radius around 10 obvious point sources in the image (only one of which is inside the SNR, at $\alpha,\delta$ = 312.56,+29.37 in the southern breakout).
We subtracted the large-scale background estimated from Sextractor \citep{Sextractor_1996}, at a scale (defined by the BACK\_SIZE parameter) set to $1\fdg5$. Then we applied adaptive smoothing using the XMM SAS task {\it asmooth} so that the signal-to-noise ratio in each pixel is at least 5$\sigma$ (the inner areas have only a few counts per pixel). The point sources that we removed were filled by this procedure because we entered the mask as an exposure map.
  None of these steps is critical to the resulting $\gamma$-ray fit. Finally, we set to 0 signal outside a circle of 1.5$\degr$ radius, with rectangular extensions covering the southern outbreak.
The 68\% angular resolution of the resulting image shown in Figure~\ref{fig:multilambda} (top left) is approximately 0.03$\degr$ (estimated from the point sources). This is much better than the $\gamma$-ray angular resolution.

In the soft X-ray band, interstellar absorption along the line of sight can significantly reduce the emitted X-ray flux. This can affect the morphology of the observed emission if absorption varies strongly across the large angular extent of this source. 
In order to estimate these variations across the SNR region, we used data from the atomic hydrogen survey HI4PI \citep{HI4PI_2016}. To focus on the foreground gas, we integrated over velocities from 0 to 10 km s$^{-1}$ (local standard of rest) in order to match the absorption value measured in 
X-rays in the interior of the remnant \citep[N$_{\rm H}$ $\sim$ 3 $\times 10^{20}$ cm$^{-2}$,][]{Uchida_2009}.
In this velocity-integrated N$_{\rm H}$ map, we observe a gradient of column density toward the Galactic plane from 3 $\times 10^{20}$ cm$^{-2}$ to 6 $\times 10^{20}$ cm$^{-2}$ from the eastern to the western bright edges of the SNR.

Assuming an average plasma temperature of 0.3 keV \citep{Katsuda_2008,Uchida_2009}, the \textit{ROSAT}/PSPC effective area and using the count rate simulator WebPIMMS \footnote{\url{https://heasarc.gsfc.nasa.gov/cgi-bin/Tools/w3pimms/w3pimms.pl}}, the count rate in the 0.1 -- 2.4 keV band varies by about 20$\%$ for the aforementioned N$_{H}$ values. We consider this effect to be negligible for our $\gamma$-ray study and did not attempt to correct for absorption effects in the X-ray map.

\subsection{Ultraviolet band}
\label{obs:UV}

The UV emission is a good tracer of the radiative shocks developing in interstellar clouds with densities of several cm$^{-3}$ (about ten times denser than the gas that is observed in X-rays). In order to cover the full Cygnus Loop, we started from the \textit{GALEX} mosaic\footnote{\url{http://www.galex.caltech.edu/media/glx2012-01r_img01.html}} kindly provided in FITS form by M. Seibert. This image was built at 3$\arcsec$ resolution from the ner-UV (NUV) images (1771 - 2831 $\AA$). The main lines from radiative shocks in this band are [C III] $\lambda1909$, [C II] $\lambda2326$ and [O II] $\lambda2471 \; \AA$.

The main difficulty with the UV mosaic is that it is dominated by point sources (and secondary reflections, so-called "smoke rings",  next to the bright ones). Therefore it cannot be used directly as a template. We applied Sextractor in two passes: a first pass with large BACK\_SIZE=128 (6$\arcmin$) to detect bright sources everywhere, and a second pass (meant to detect faint sources while avoiding removal of pieces of filaments) with smaller BACK\_SIZE=32 (1.5$\arcmin$) followed by a selection (based on source angular size, flags, and flux/background ratio) requiring that a detection looks like a point source. We generated circular regions  excluding the entire regions where sources increase the background visually (radius proportional to flux to the power of 0.3), adapted a few by eye, and added 80 regions around the secondary reflections. This resulted in about 10,000 excluded regions in total.

    After this, we rebinned the masked image and the mask into 30$\arcsec$ pixels (we do not need better angular resolution to fit the $\gamma$-rays). We smoothed the image locally around the zero values in the rebinned mask (where bright stars were) and divided the smoothed image by the smoothed mask to recover a flat exposure.
    The last stage (large-scale background subtraction, adaptive smoothing, and clipping) was the same as in \ref{obs:Xray}, except that BACK\_SIZE was set to 32$\arcmin$ because there are no large-scale features in the UV image.
    The resulting image is shown in Figure~\ref{fig:multilambda} (top right).

\subsection{Optical band}
\label{obs:Opt}

The optical band also traces radiative shocks. Because the lines in this wavelength range are not the same as in the UV, its sensitivity to different shock speeds and different ages can be slightly different.
Again because of the angular size of the Cygnus Loop, a sky survey is better. We therefore used the Digital Sky Survey 2 (DSS2) images in the red band, which covers 6000 to 7000 $\AA$, including [O I] $\lambda6300$, H$\alpha \;\lambda6536$, [N II] $\lambda6584,$ and [S II] $\lambda6717-6730 \; \AA$.
We obtained the data from the STScI server\footnote{\url{http://archive.stsci.edu/dss}}, forcing plate XP463 in order to preserve a uniform background (no automatic jump to plate XP464). We extracted $3 \times 4$ $60 \times 60\arcmin$ images at 1$\arcsec$ resolution separated by 55$\arcmin$, which provided coverage of the full SNR with 2.5$\arcmin$ overlap between images.

The principle of source detection and exclusion was the same as in the UV, with the additional difficulty that the bright stars saturate the plates and look broader than the faint stars. We therefore used three different Sextractor settings, which reach deeper while using smaller BACK\_SIZE and DETECT\_MINAREA. The third run (for the faint sources) also required that the detections looked like point sources.
About 5,000 regions were excluded from each image on average. The radius of the circles was set to twice the source full width at half maximum (FWHM) reported by Sextractor for the bright sources, and 1.5 FWHM for the medium and faint sources.
We rebinned the images to 3$\arcsec$ (aligned with the UV image as much as possible) to avoid needlessly large files, built the mosaic of $3 \times 4$ original images to cover the entire Cygnus Loop, and then rebinned to the final 30$\arcsec$ pixels.
The last stages (point-source filling, large-scale background subtraction, adaptive smoothing, and clipping) were the same as in  \ref{obs:UV}.
The resulting image is shown in Figure~\ref{fig:multilambda} (bottom left).
From the $\gamma$-ray point of view, the optical image is very similar to the UV. However, the brightest filaments approach the saturation level in the DSS2 images.

\subsection{Radio band}
\label{obs:Radio}

The radio band (synchrotron emission) traces a combination of low-energy electrons and magnetic field.
In order to ensure that the large-scale signal was preserved, we used the single-dish images reported in \citet{Uyaniker_2004} from the \textit{Effelsberg} 100 m telescope at 21 and 11 cm, and \citet{Sun_2006} from the \textit{Urumqi} 25 m telescope at 6 cm. The respective half-power beam widths were 9.4$\arcmin$, 4.3$\arcmin$ and 9.5$\arcmin$. This is better than the $\gamma$-ray resolution, but not good enough to extract point sources self-consistently as in the other wavebands.
Instead, we used the NVSS catalog \citep{Condon_1998} that was obtained at higher resolution with the \textit{VLA} to select 40 point sources brighter than 100 mJy and with intrinsic sizes smaller than 100$\arcsec$ in the field of view. We excluded disks with a radius 0.1$\degr$ at 11 cm (0.15$\degr$ at 6 and 21 cm), scaled by $F_{\rm Jy}^{0.3}$ as we did in the UV ($F_{\rm Jy}$ is the source flux in Jy). We refilled them by smoothing as in \ref{obs:UV}.
 Similarly to the X-rays, the Cygnus Loop is bright enough in the radio for none of these steps to be critical for the resulting $\gamma$-ray fit.
    The map with the best angular resolution (at 11 cm) is shown in Figure~\ref{fig:multilambda} (bottom right).

\section{Analysis}
\label{section:Analysis}

\begin{figure*}
   \centering
            \includegraphics[width=0.99\textwidth]{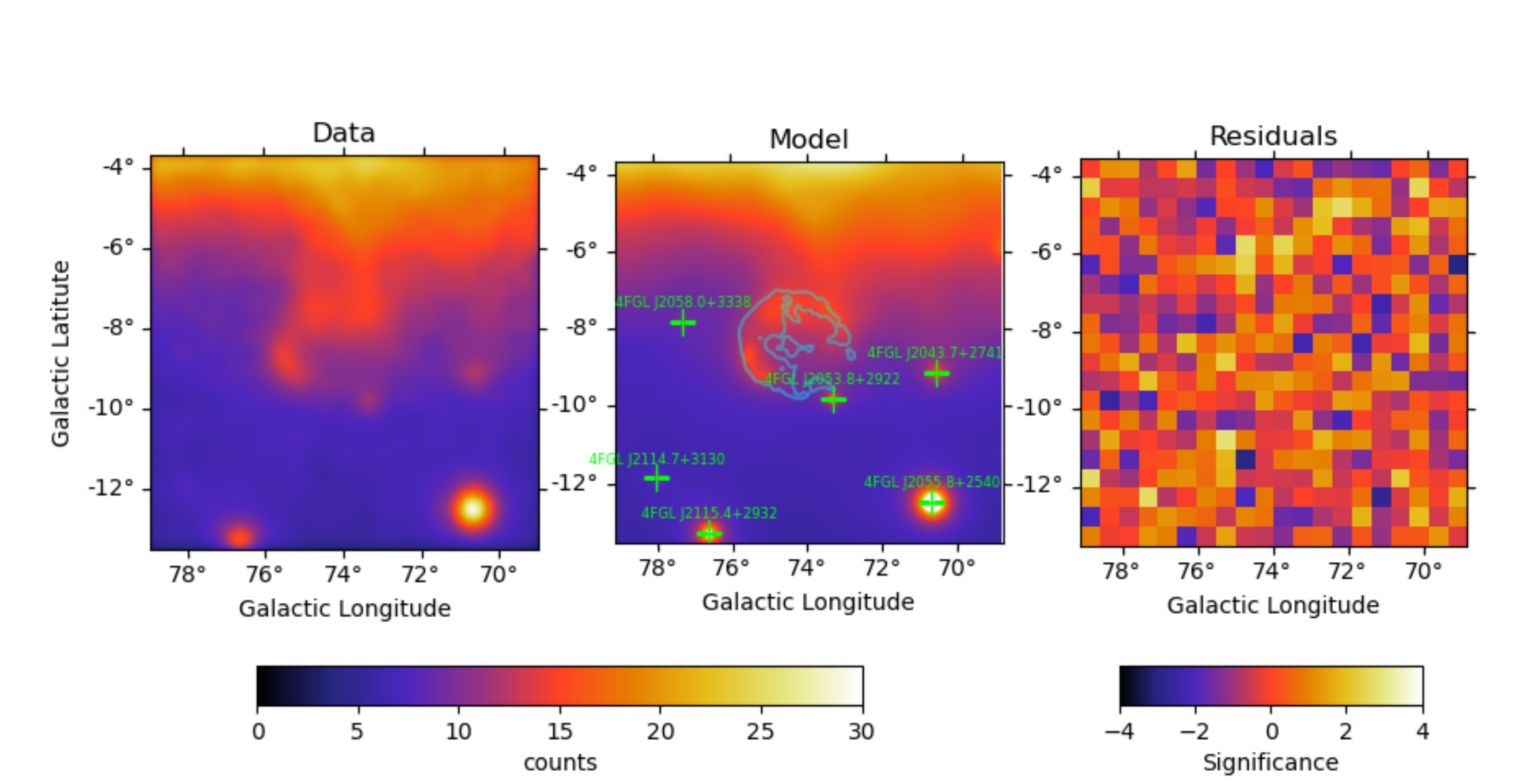}
   \caption{Maps of a $10^{\circ} \times10^{\circ}$ region around the Cygnus Loop. \textit{Left:} Count map (pixel size of $0.05^{\circ}$) smoothed with a Gaussian kernel of $0.2^{\circ}$ from 0.1 to 100 GeV. \textit{Center:} Count map expected from the X-ray+UV model (same spatial binning as the left map).  Green crosses indicate the positions of $\gamma$-ray sources listed in the 4FGL catalogue. The cyan line is the contour ($10\%$ of the maximum) of the \textit{ROSAT} X-ray template of the Cygnus Loop (see \ref{obs:Xray}). \textit{Right:} Residual count map from the X-ray+UV template model (pixel size of $0.5^{\circ}$).}
   \label{fig:excess}
   \end{figure*}
   
\begin{figure*}
   \centering
            {\includegraphics[viewport=4 38 363 356,clip,width=0.45\textwidth]{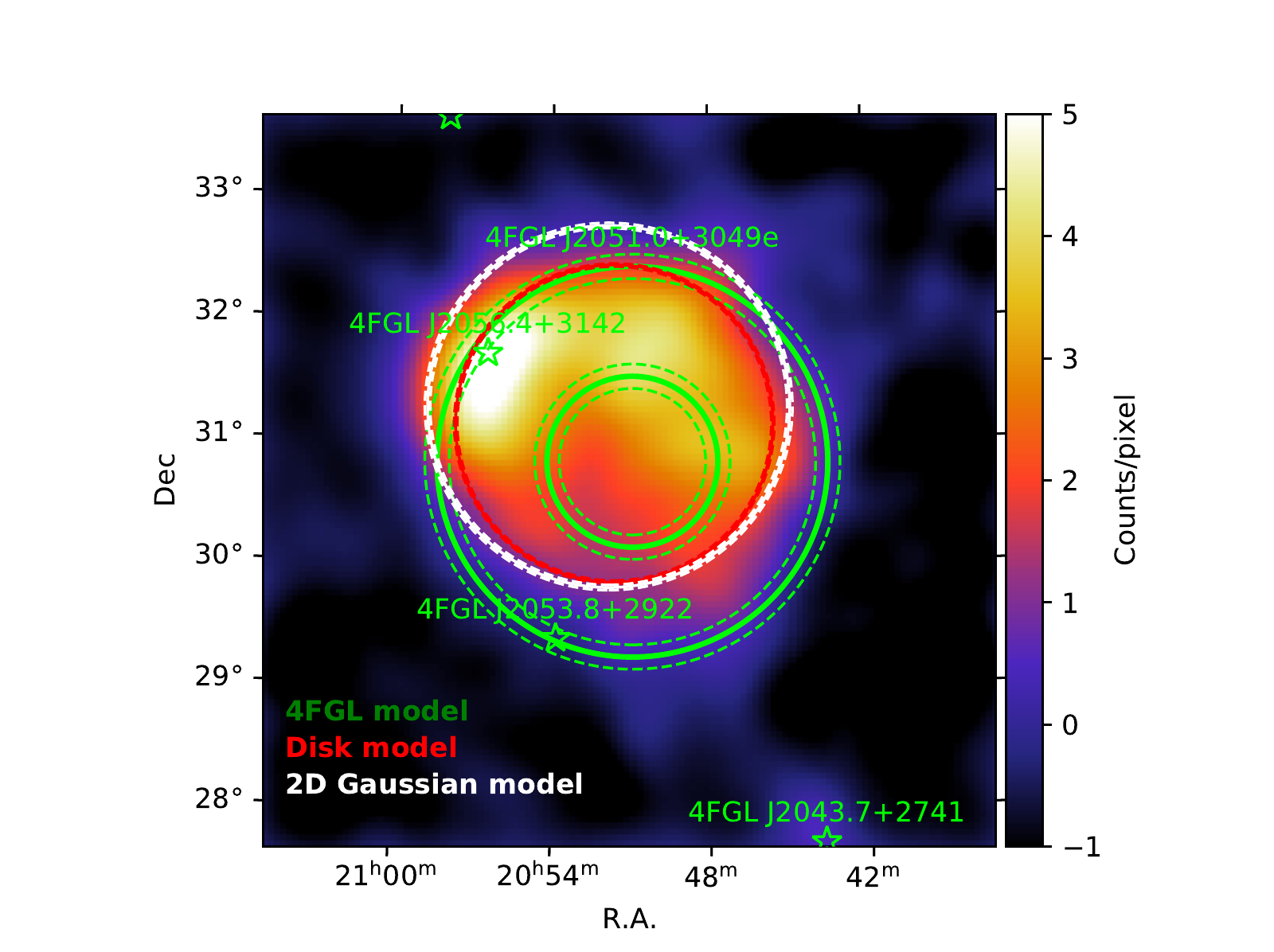}} \quad
            {\includegraphics[viewport=71 38 430 356,clip,width=0.45\textwidth]{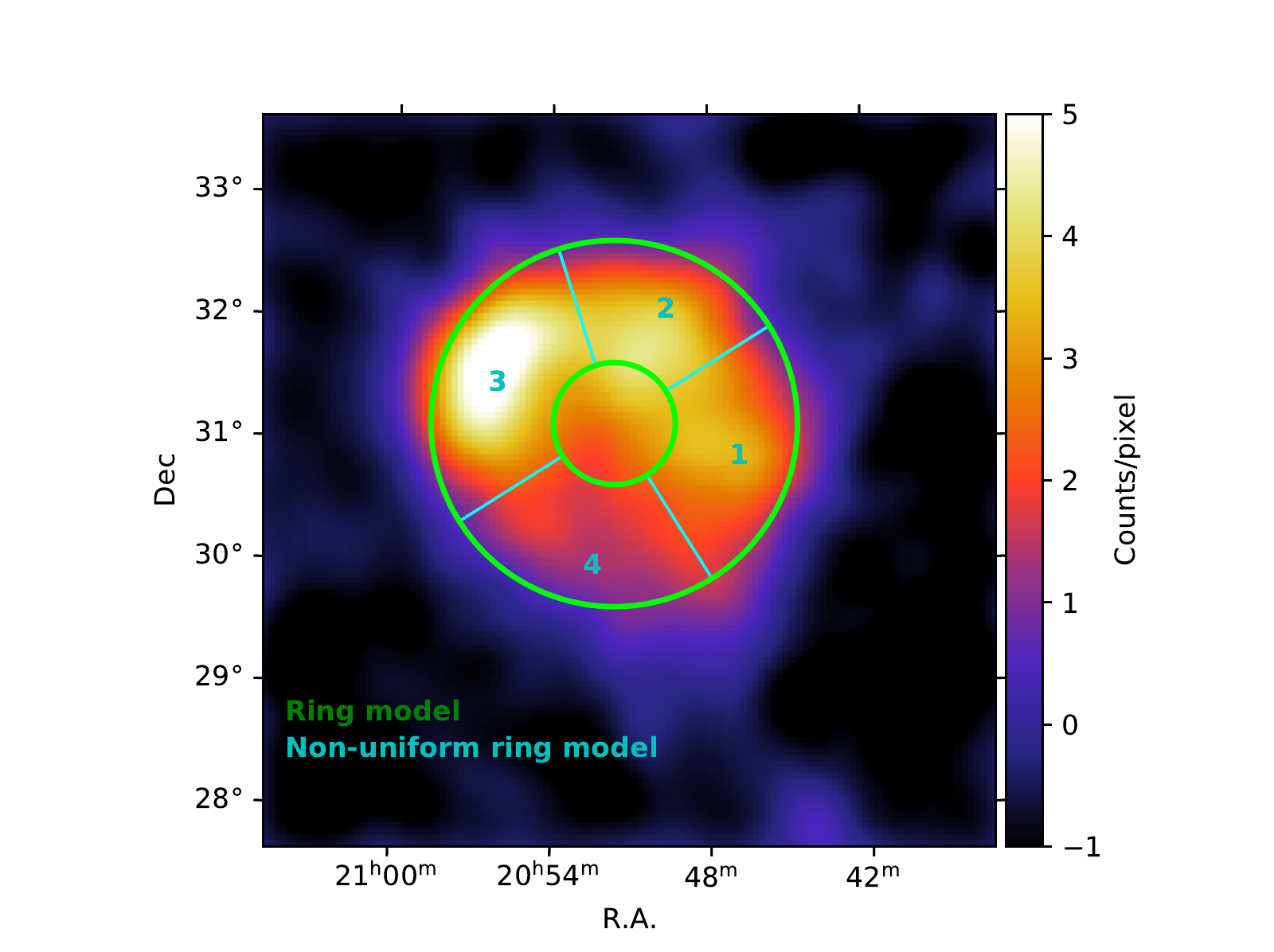}} \\
            {\includegraphics[viewport=4 1 363 328,clip,width=0.45\textwidth]{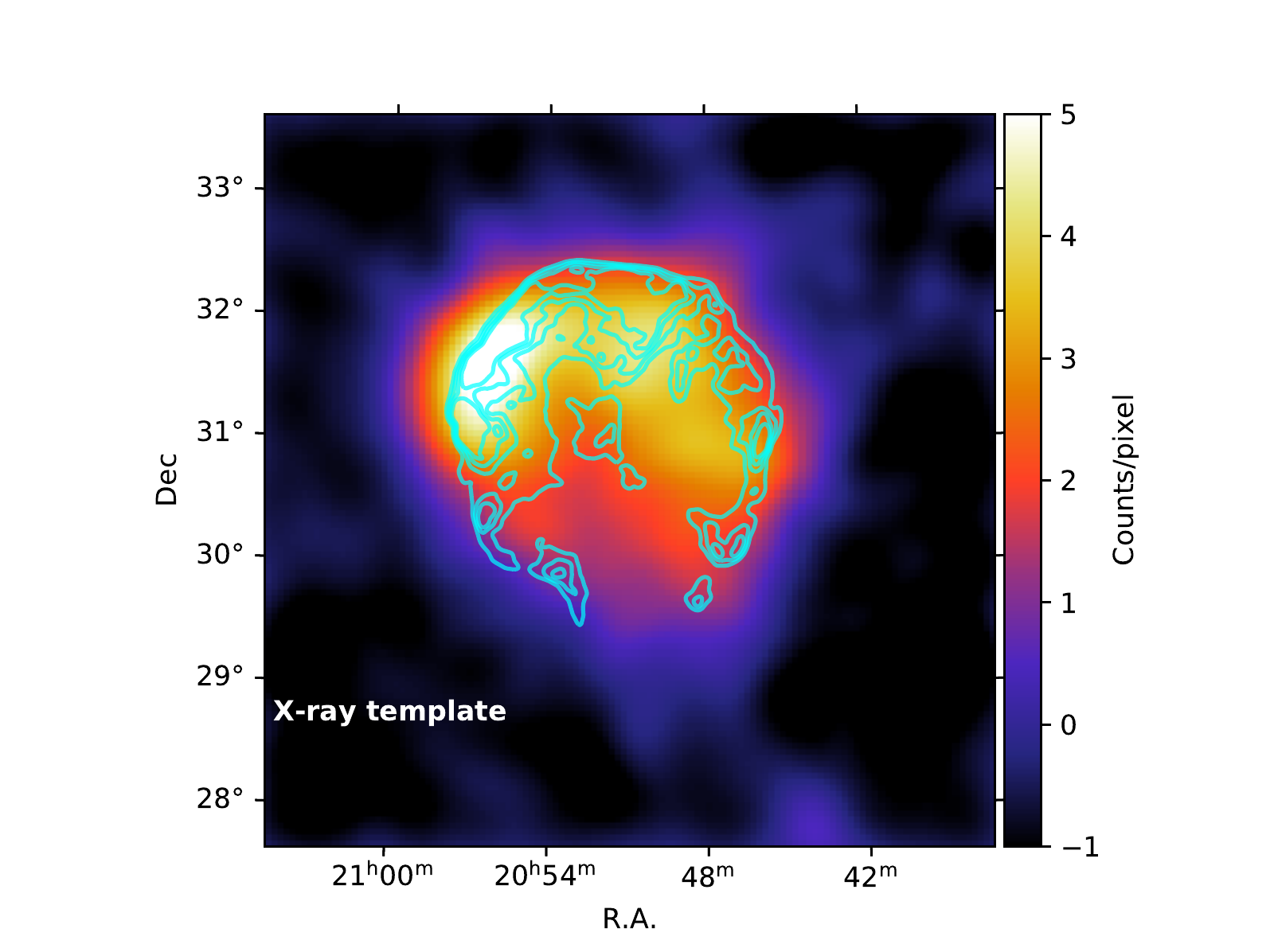}} \quad
            {\includegraphics[viewport=71 1 430 328,clip,width=0.45\textwidth]{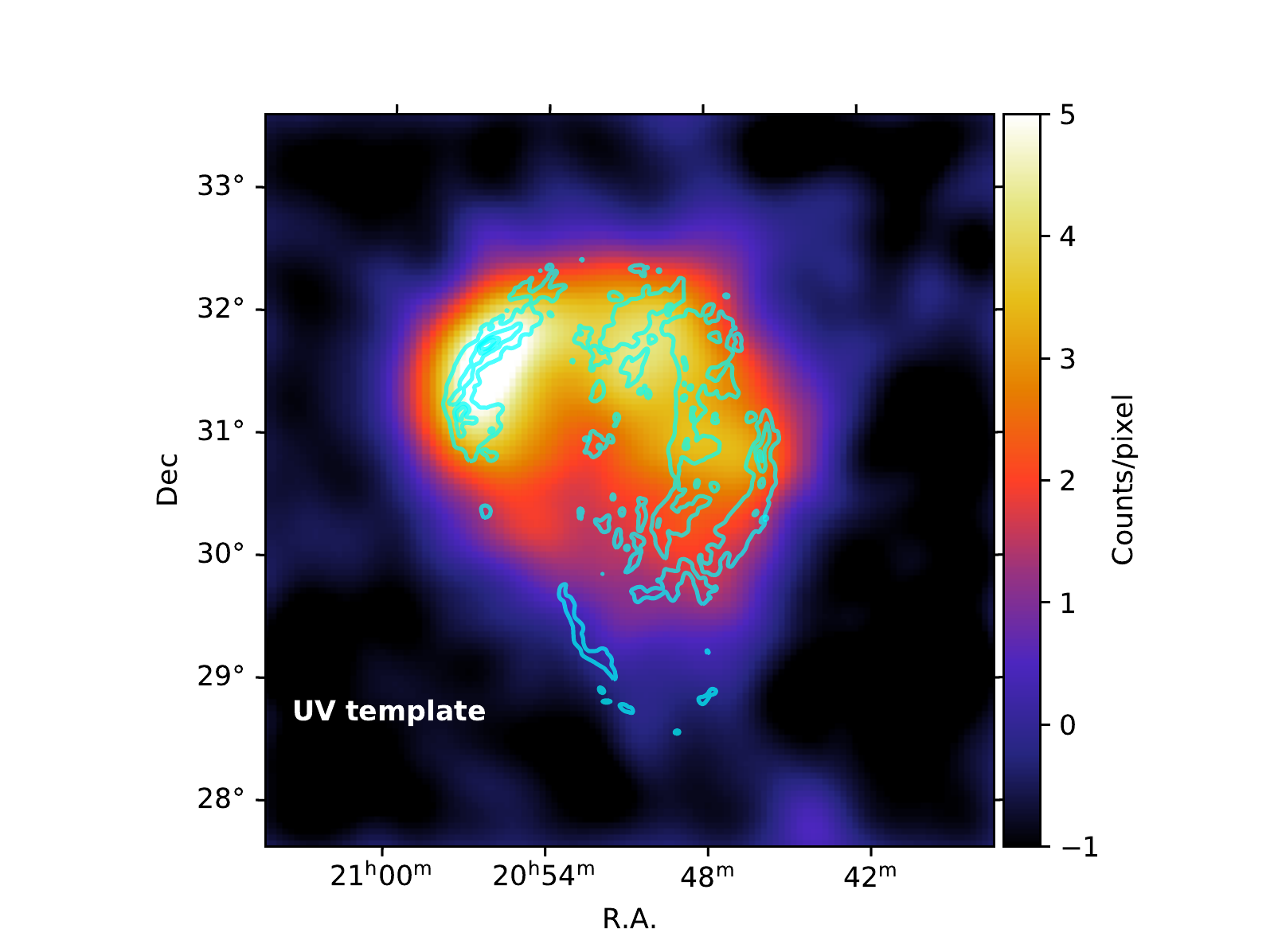}} \\
            
   \caption{Residual count map in a $6^{\circ} \times6^{\circ}$ (pixel size of $0.05^{\circ}$) region around the Cygnus Loop (smoothed with a Gaussian kernel of $0.2^{\circ}$) obtained from 0.1 to 100 GeV. The Cygnus Loop was not included in the model (\textit{null hypothesis}). Different templates are overlaid. \textit{Top left:} Best-fit disk and Gaussian models represented in red and white, respectively. The 4FGL point sources are shown as green stars. The ring model (4FGL J2051.0+3049e) introduced by \citet{Katagiri_2011} is shown in green. \textit{Top right:} Best-fit ring model (green). The blue lines define the four sections used in Tables~\ref{tab:spatial} and \ref{tab:panda}. \textit{Bottom left:} Contours of the \textit{ROSAT} X-ray template (cyan, see \ref{obs:Xray}). The templates were smoothed with a Gaussian kernel of $\sigma = 0.2^{\circ}$ to make the contours more regular. Contours for the X-ray template are at $30\%, 20\%, 10\%,$ and $1\%$ of the maximum. \textit{Bottom right:} Contours for the \textit{GALEX} UV template (cyan, see \ref{obs:UV}) are at $40\%, 25\%, 15\%,$ and $2\%$ of the maximum.}
   \label{fig:contour}
   \end{figure*}
   
The 4FGL catalog records three sources around the position of the Cygnus Loop: The extended ring (4FGL J2051.0+3049e) introduced by \citet{Katagiri_2011}, a point source in the eastern part of the ring (4FGL J2056.4+3142), and a point source in the southern part of the ring (4FGL J2053.8+2922). They are all described by LP spectra (see the left and middle panel in Figure~\ref{fig:excess}). While the former two sources are associated with the Cygnus Loop, the latter is associated with an AGN \citep[RX J2053.8+2923,][]{Brinkmann_1997} in the 4FGL catalog.

We performed the morphological analysis from $0.1$ GeV to $100$ GeV. The free parameters in the model were the normalizations of the sources located closer than $6^{\circ}$ to the ROI center, of the Galactic and isotropic diffuse emissions, and the spectral parameters of the Cygnus Loop and 4FGL J2053.8+2922.
The nearest bright sources are PSR J2028+3332 and PSR J2055+2539, which are stable sources farther away than 5$\degr$ from the ROI center.

\subsection{Geometrical models}

\begin{table*}
\caption{Best-fit spatial properties and test statistics of the Cygnus Loop for different morphological models compared with the null hypothesis of no $\gamma$-ray emission associated with the Cygnus Loop (0.1–100 GeV).}
\label{tab:spatial}
\begin{tabular}{lccccc} 
\hline 
Model & RA$_{J2000} (^{\circ})$ & DEC$_{J2000} (^{\circ})$ & $r - \sigma (^{\circ})$ & Test Statistic \tablefootmark{a} & Additional DoF\\ 
\hline 
Null hypothesis \tablefootmark{b}   &                      &                     &                                  &    0 & 0 \\
Uniform Disk                        & $312.905 \pm 0.016$  & $31.130 \pm 0.018$  & $1.300^{+0.011}_{-0.014}$        & 6120 & 6 \\
Gaussian 2D                         & $312.959 \pm 0.022$  & $31.269 \pm 0.021$  & $1.483^{+0.021}_{-0.021}$        & 5874 & 6 \\
Ring                                &      $312.905$       &      $31.130$       & $r_{max} = 1.50^{+0.01}_{-0.02}$, $r_{min} = 0.50^{+0.04}_{-0.07}$ & 6172 & 7 \\
Non-uniform ring \tablefootmark{c}  &      $312.905$       &      $31.130$       & $r_{max} = 1.50$, $r_{min} = 0.50$ & 6442 & 16\\
X-ray ($0.1-2$ keV)\tablefootmark{d}        & & & & 6446 & 3 \\
Optical ($1.7-2$ eV) \tablefootmark{e}      & & & & 6584 & 3 \\
UV ($4-9$ eV) \tablefootmark{f}             & & & & 6650 & 3 \\
Radio 21cm ($1427$ MHz) \tablefootmark{g}   & & & & 4986 & 3 \\
Radio 11cm ($2725$ MHz) \tablefootmark{g}   & & & & 4878 & 3 \\
Radio 6cm ($4996$ MHz)  \tablefootmark{h}   & & & & 4750 & 3 \\
X-ray + UV                                  & & & & 6870 & 6 \\
\hline
\end{tabular}
\tablefoot{
Spectral model: LogParabola.}
\tablefoottext{a}{$ -2 \log (\mathcal{L}_0 /\mathcal{L})$, where $\mathcal{L}$ and $\mathcal{L}_0$ are the maximum likelihoods for the model with or without the source component, respectively.} \\
\tablefoottext{b}{Background only (no model for the Cygnus Loop).} \\
\tablefoottext{c}{A ring (same radius as the ring model) divided into four regions as shown in Figure~\ref{fig:contour} and allowing an independent normalization and spectral index for the four portions of the ring.} \\
\tablefoottext{d}{The X-ray template is based on data from \textit{ROSAT} (see \ref{obs:Xray}).} \\
\tablefoottext{e}{The optical template is based on data from \textit{DSS2} (see \ref{obs:Opt}).} \\
\tablefoottext{f}{The UV template is based on data from \textit{GALEX} (see \ref{obs:UV}).} \\
\tablefoottext{g}{Template based on data from the \textit{Effelsberg 100 m Radio Telescope} (see \ref{obs:Radio}).} \\
\tablefoottext{h}{Template based on data from the \textit{Urumqi 25 m Telescope} (see \ref{obs:Radio}).}
\end{table*}

To perform the morphological analysis, we considered the model without emission from the Cygnus Loop (we removed sources 4FGL J2051.0+3049e and 4FGL J2056.4+3142) as our \textit{\textup{null hypothesis}}, which has a maximum likelihood $\mathcal{L}_0$. Figure~\ref{fig:contour} shows the excess map of a $6^{\circ} \times6^{\circ}$ region, with a pixel size of $0.05^{\circ}$ and centred on the Cygnus Loop position, obtained using our \textit{\textup{null hypothesis}} as a model. Then, we tested alternative models by adding spatial templates and/or by varying the parameters of the models, and we computed the corresponding maximum likelihood $\mathcal{L}_{mod}$. The fit improvement is quantified by the test statistic \citep{Mattox_1996},
\begin{equation}
    TS = 2 \times (\log \mathcal{L}_{mod} - \log \mathcal{L}_0)
\label{eq:llratio}
,\end{equation}
which in the absence of a real source, follows a $\chi^2$ distribution with $k$ degrees of freedom corresponding to the additional free parameters between the models and the \textit{\textup{null hypothesis}}.

We tested several spatial models to describe the emission from the Cygnus Loop, assuming an LP spectrum with all parameters free. First, we started with three geometrical models: a uniform disk, a 2D symmetric Gaussian, and a uniform ring. We report the best-fit positions and extensions we obtained, with the associated TS values for these models, in Table~\ref{tab:spatial}. The Akaike information criterion (AIC;~\citealt{Akaike_1974}) was adopted to compare the different geometrical models, where the AIC values are computed as $AIC = 2k - 2\log \mathcal{L}$ ($k$ is the number of estimated free parameters in the model). The result in Table~\ref{tab:spatial} shows an obvious improvement when the disk model rather than the Gaussian model is used ($\Delta AIC = AIC_{Gauss} - AIC_{Disk} = 246 $). To explore the uniform ring template, we defined a 2D ring with a morphology defined by a FITS template. We kept the ring centered at the best-fit position of the disk model, and we varied inner and outer radii and evaluated the maximum likelihood values. We explored values of inner radius in the range of $r_{min} = 0.2^{\circ} - 0.6^{\circ}$ and of the outer radius $r_{max} = 1.5^{\circ} - 1.7^{\circ}$. In Table~\ref{tab:spatial} we report the best model. The $\Delta AIC$ value for the ring with respect to the disk shape is $50$: the ring is clearly favored.

Figure~\ref{fig:contour} shows that the emission along the remnant is not very uniform. We therefore also searched for a possible spectral variation in the $\gamma$-ray emission along the Cygnus Loop. To this end, we divided our best-fit ring into four sections, as shown in the top right panel in Figure~\ref{fig:contour}. We independently fit the four sections leaving the normalizations, $\alpha$ and $\beta$ parameters, as free parameters. This leads to a higher TS value than the uniform ring because the nonuniform ring can handle the differences along the remnant better. The results, shown in Table~\ref{tab:panda}, indicate that there are no significant differences in the spectral indices along the Cygnus Loop. The $\gamma$-ray emission is fainter in the southern region (region 4) and brighter in the northeast (region 3).

\subsection{Correlation with other wavelengths}
We further investigated the Cygnus Loop morphology by evaluating the correlation with emission at other wavelengths: X-rays, UV (see bottom panels in Figure~\ref{fig:contour}), and radio continuum. We used the images at these wavelengths as spatial templates to fit the $\gamma$-ray emission, assuming an LP spectrum. The TS values for the X-ray, optical, and UV templates show a large improvement compared to all the other models. The UV template is best, but even the X-ray template is favored compared to the nonuniform ring ($\Delta AIC$ = 38) because it has far fewer degrees of freedom. The optical template is somewhat worse than the UV, but this is probably due to the saturation of the DSS2 images (\ref{obs:Opt}). In contrast, the radio templates have lower TS values because of their bright emission in the southern region of the remnant, where the $\gamma$-ray emission is fainter. This difference between the radio emission and the other wavebands has been explained by the existence of a separate SNR interacting with the Cygnus Loop~\citep{Uyaniker_2002}, although a recent multiwavelength analysis~\citep{Fesen_2018} makes this interpretation controversial.

The X-ray distribution follows the rims of the remnant, and its correlation with the $\gamma$-ray emission seems to suggest that the high-energy particles may be located in the forward-shock region. The UV template instead traces the radiative shocks in the remnant, and its filamentary structures are correlated with the central and west regions of the remnant.
The residual map after fitting the X-ray (UV) template shows significant emission correlated with the UV (X-ray) template. We therefore tested a two-component model, including both the X-ray and UV maps. The sharp increase in the TS parameter (by $\sim$ 200; see Table~\ref{tab:spatial}) together with the residuals within 4$\sigma$ (see the right panel of Figure~\ref{fig:excess}) indicates that the X-ray+UV templates fit the $\gamma$-ray morphology adequately. The residuals are normally distributed around a mean value of 0.05$\sigma$ with a standard deviation $\sigma_{\rm tot} = 1.106\sigma$, implying a systematic contribution of $\sigma_{\rm syst} = \sqrt{\sigma_{\rm tot}^2 - 1} = 0.47\sigma$, which is lower than the statistical contribution.

\begin{table}
\caption{Test statistic and best-fit spectral indices for the four sections of the ring as shown in Figure~\ref{fig:contour} using a log parabola model.}.
\label{tab:panda}
\resizebox{\columnwidth}{!}{
\begin{tabular}{lcrrc} 
\hline 
\multirow{2}*{Region} & \multirow{2}*{TS} & \multirow{2}*{$\alpha$} & \multirow{2}*{$\beta$}  & Normalization\tablefootmark{a}\\ 
                      &                   &                         &                         & ($10^{-12}$ MeV$^{-1}$ cm$^{-2}$ s$^{-1}$)\\ 
\hline 
1  & 916  & $2.10 \pm 0.04$  & $0.13 \pm 0.02$ & $5.74 \pm 0.26$\\
2  & 586  & $2.01 \pm 0.05$  & $0.20 \pm 0.01$ & $4.23 \pm 0.22$\\ 
3  & 2194 & $2.07 \pm 0.03$  & $0.17 \pm 0.06$ & $8.50 \pm 0.22$\\
4  & 206  & $1.91 \pm 0.08$  & $0.15 \pm 0.03$ & $2.21 \pm 0.22$\\
\hline

\end{tabular}
}
\tablefoot{}
\tablefoottext{a}{The normalization is computed at 837 MeV, following \citet{Katagiri_2011}}
\end{table}
  
\subsection{Spectral analysis}
\label{section:spectral}

\begin{table*}
\caption{Test statistics and parameters for various spectral models (0.1 – 100 GeV) using as spatial model the UV map.}
\label{tab:spectral}
\begin{tabular}{lcccc} 
\hline 
\multirow{2}*{Model} & \multirow{2}*{TS \tablefootmark{a}} & \multirow{2}*{Spectral Model\tablefootmark{b}} & \multirow{2}*{Spectral Parameters} & Energy flux (0.1 -- 100 GeV)\\
                     &                                     &                               &                                    & ($10^{-5}$ MeV cm$^{-2}$ s$^{-1}$) \\
\hline 
Power Law (PL)                           & 0 &  $E^{-p}$   &$p = 2.14 \pm 0.01$ & $7.15 \pm 0.12$ \\
\hline
\multirow{2}*{Exp. cut-off PL}     & \multirow{2}*{324} &  \multirow{2}*{$\Big(\frac{E}{E_0}\Big)^{-\gamma}\exp{-\Big(\frac{E}{E_{cut}}\Big)}$} & $\gamma = 1.77 \pm 0.03$ & \multirow{2}*{$5.31 \pm 0.10$} \\
                                   &                    & &$E_{cut} = 4.1 \pm      0.4$ GeV & \\
\hline
\multirow{2}*{LogParabola}         & \multirow{2}*{418} &  \multirow{2}*{$\Big(\frac{E}{E_0}\Big)^{-\alpha-\beta\log(E/E_0)}$} & $\alpha = 2.134 \pm 0.016$ & \multirow{2}*{$5.44 \pm 0.10$}\\
                                   &                    &  & $\beta = 0.212 \pm 0.013$ & \\
\hline
\multirow{3}*{PLSuperExpCutoff4}   & \multirow{3}*{428} & \multirow{3}*{$\left(\frac{E}{E_0}\right)^{-\gamma_0 + d/b}\exp\left[\frac{d}{b^2}\left(1-\left(\frac{E}{E_0}\right)^b\right)\right]$} & $\gamma_0 = 2.140 \pm 0.024$ & \multirow{3}*{$5.62 \pm 0.12$} \\
                                   &                        & & $d = 0.350 \pm 0.023$ & \\
                                   &                        & & $b = -0.40 \pm 0.12$ & \\
\hline
\multirow{4}*{Smoothly broken PL}           & \multirow{4}*{430} &  \multirow{4}*{$\Big(\frac{E}{E_0}\Big)^{-p_1}\left[1+\left(\frac{E}{E_b}\right)^{\frac{p_2-p_1}{b}}\right]^{-b}$} & $p_1 = 2.68 \pm 0.07$ & \multirow{4}*{$5.82 \pm 0.11$} \\
                                            &                    & & $p_2 = 1.12 \pm 0.12$ & \\
                                            &                    & & $E_b = 620 \pm 130$ MeV & \\
                                            &                    & & $b = 0.99 \pm 0.04$ & \\                         
\hline
\end{tabular}
\tablefoot{}
\tablefoottext{a}{$ -2 \ln(\mathcal{L}_0/\mathcal{L})$, where $\mathcal{L}$ and $\mathcal{L}_0$ are the maximum likelihood values for the model under consideration and the power-law model, respectively.} \\
\tablefoottext{b}{The reference energy $E_0$ is set to 1 GeV for all the models.} \\
\end{table*}

\begin{table}
\caption{Best-fit spectral parameters for the X-ray+UV model using a LogParabola model.}
\label{tab:X+UV}
\begin{tabular}{lccc} 
\hline 
\multirow{3}*{Region} & \multirow{3}*{$\alpha$} & \multirow{3}*{$\beta$} & Energy flux \\
                      &                         &                        & (0.1 -- 100 GeV) \\
                      &                         &                        & ($10^{-5}$ MeV cm$^{-2}$ s$^{-1}$) \\
\hline 
X-ray  & $2.17 \pm 0.07$ & $0.37 \pm 0.04$ & $2.07 \pm 0.16$\\
UV     & $1.97 \pm 0.04$ & $0.14 \pm 0.02$ & $3.48 \pm 0.19$\\ 
\hline
\end{tabular}
\end{table}

Using the UV template as a morphological model for the remnant, we investigated the spectral shape of the Cygnus Loop as a whole. We compared the likelihood values of the spectral fit for a power law with other spectral functions over the entire considered energy range. TS values and best-fit parameters are reported in Table~\ref{tab:spectral}. A curved spectrum is clearly preferred over a power-law spectrum. The exponentially cutoff power law is not a good model (the spectrum does not fall off exponentially toward high energies). A simple symmetric log-normal (LogParabola) model fits the data quite well. The PLSuperExpCutoff4 model in the \texttt{Fermitools} has a superexponential index $b<0$ (\textup{i.e.}, with a subexponential fall-off toward low energies and power-law decrease toward high energies). It increases TS by 10, corresponding to an improvement of slightly more than 3$\sigma$ with respect to LogParabola (which corresponds to $b$ = 0, with $d = 2\beta$; see the parameter definitions in Table~\ref{tab:spectral}).
The smoothly broken power-law model (with one more parameter) does not improve the fit. Considering the three possible spectral models, the integrated energy flux of the Cygnus Loop using the UV template in the energy band of 0.1–100 GeV is $5.6 \pm 0.2 \times 10^{-5}$ MeV cm$^{-2}$ s$^{-1}$.

\begin{figure}
   \centering
   {\includegraphics[width=\hsize]{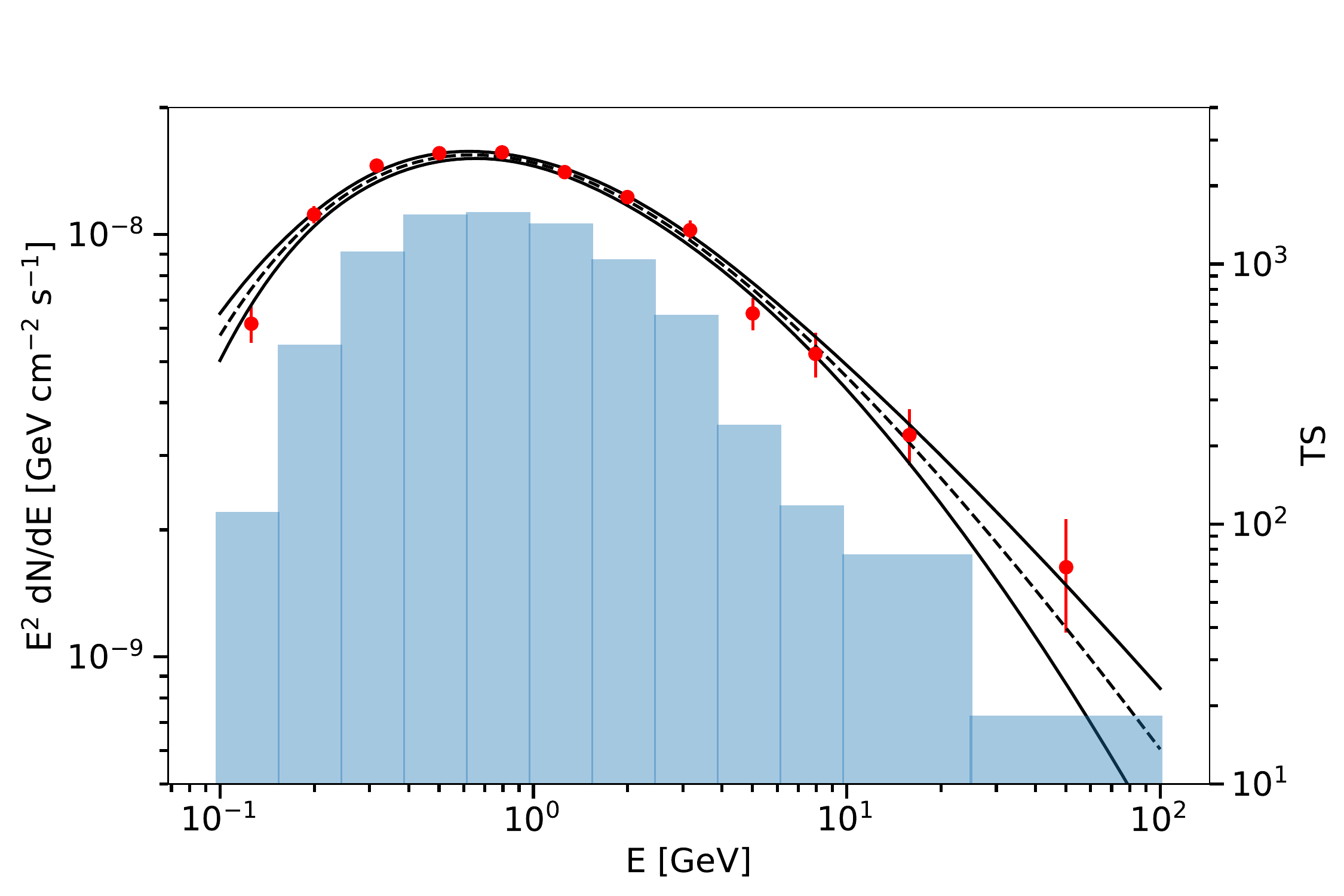}} \quad
   {\includegraphics[width=\hsize]{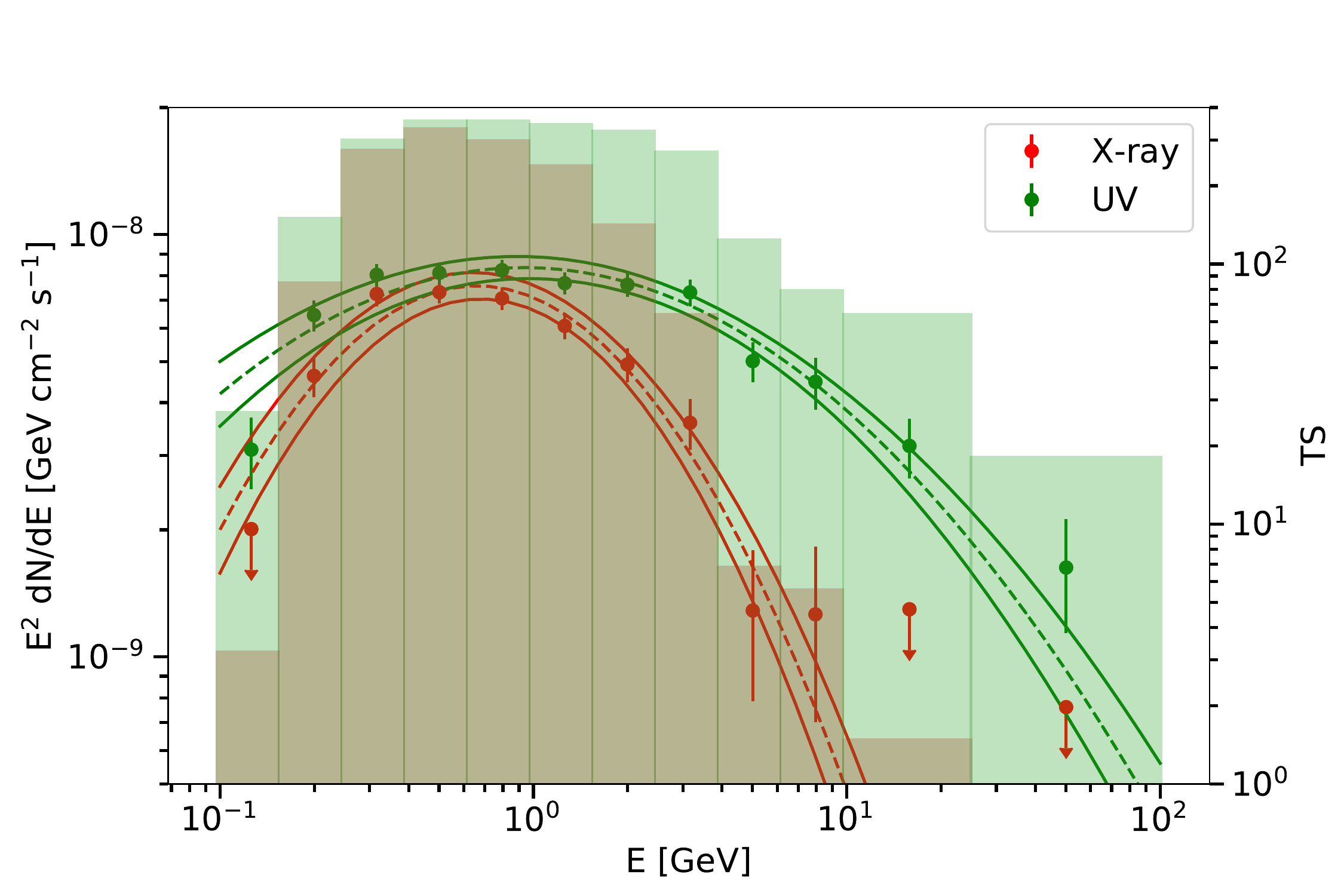}} \\
      \caption{Spectral energy distribution of the $\gamma$-ray emission measured with the LAT for the Cygnus Loop. The filled histograms show the TS values in each energy bin. Vertical bars show $1\sigma$ statistical errors. Where the detection is not significant (TS < 4), we show upper limits at the 95\% confidence level. \textit{Upper panel:} SED extracted using the UV template. The PLSuperExpCutoff4 best-fit spectrum for the global $\gamma$-ray data (Table \ref{tab:spectral}) is plotted as the dashed black line, and its upper and lower $1\sigma$ bounds as the solid black lines. \textit{Lower panel:} Red (green) points are LAT flux points using the X-ray (UV) maps as spatial templates together. The lines are the best-fit LogParabola models (Table \ref{tab:X+UV}).}
         \label{fig:seds}
   \end{figure}

We extracted the spectral energy distribution (SED) in ten logarithmically spaced energy bands from 0.1 to 10~GeV and two broader bins above 10~GeV. In each bin, the photon index of the source was fixed to 2, and we imposed a TS threshold of 4, below which we calculated an upper limit. The upper panel of Figure~\ref{fig:seds} shows the resulting SED.

We then performed a spectral analysis using our best morphological template: the X-ray+UV template. Statistics are not enough to constrain models with more than two shape parameters such as PLSuperExpCutoff4, but we explored different combinations of power-law and LogParabola functions for each one of the two components (the X-ray and UV template). Using an LP spectral function for both the X-ray and UV templates, we obtained the highest TS value. It is increased by 61 compared to fixing the shape parameters to the best PLSuperExpCutoff4 of Table~\ref{tab:spectral} and fitting only normalizations. This implies that the spectral shapes of the X-ray and the UV components differ significantly. The results are summarized in Table~\ref{tab:X+UV}. We then extracted the SED of both components as explained previously. The lower panel of Figure~\ref{fig:seds} shows the resulting SEDs.

\subsubsection{Radio data extraction}
\label{section:radio_data}

A major difficulty when fitting the nonthermal emission of the Cygnus Loop is that the radio maps do not look like the $\gamma$-ray, X-ray or UV/optical maps (Figure~\ref{fig:multilambda}). Radio maps alone show strong emission toward the southwest, and indeed, the radio template is by far the worst fit to the $\gamma$-ray data (Table~\ref{tab:spatial}).
We are interested in the part of the radio emission that follows the other wavebands.
More precisely, because we have shown (Table~\ref{tab:spatial}) that the $\gamma$-ray data are well fit by a combination of the UV and X-ray templates, we wish to decompose the radio emission in the same way.

\begin{figure*}
\centering
\includegraphics[width=0.90\textwidth]{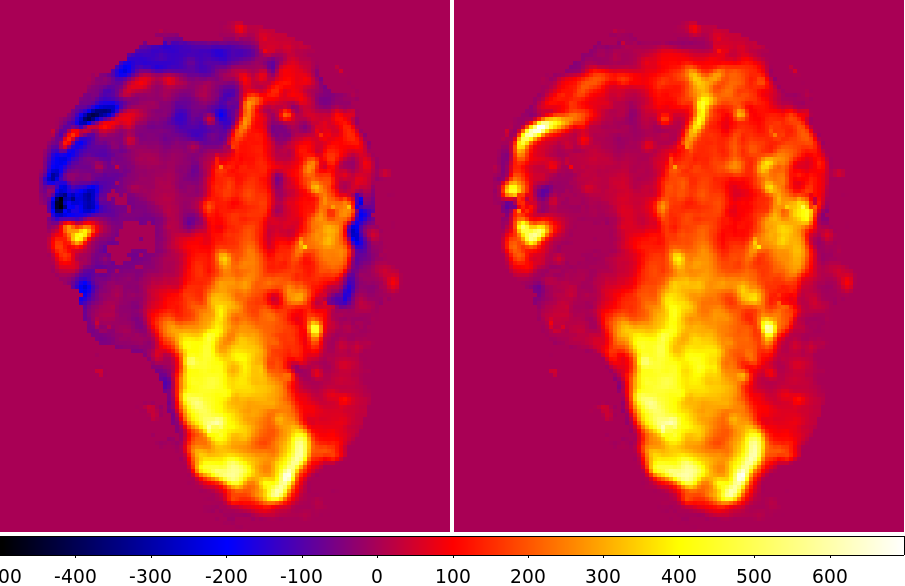}
\caption{Residuals after fitting the 11 cm Effelsberg map with a combination of the UV and X-ray templates. Left: Standard $\chi^2$ fit resulting in deep negative residuals. Right: Fit increasing the errors by a factor 10 for positive residuals, without negative residuals but higher positive residuals.}
\label{fig:radioresid}
\end{figure*}

The first step is to convolve the UV and X-ray templates into the radio PSF (different in each band). Because the X-ray angular resolution $\sigma_X$ is not negligible with respect to the radio angular resolution $\sigma_R$, this was achieved by a convolution with a Gaussian of $\sigma^2 = \sigma_R^2 - \sigma_X^2$. We primarily worked on the 11 cm map, which has the best angular resolution and signal-to-noise ratio among the three radio maps. Fitting these convolved templates to the radio maps using a standard $\chi^2$ fit results in deep negative residuals in the northeast and west (Figure~\ref{fig:radioresid} left) because the fit tries to push the UV and X-ray templates as high as possible to account for the radio structure. We instead searched for a decomposition that would leave only positive residuals, corresponding to the part of the radio emission that is uncorrelated with the UV and the X-ray emitting regions. In order to achieve this in a simple way, we increased the errors in the $\chi^2$ formula by a factor $R$ wherever the residuals are positive. Figure~\ref{fig:radioresid} (right) shows that for $R$ = 10,  no negative residuals are left. The fraction of the total radio flux in the residuals is 40\% for $R$ = 1, and this increases to 78\% for $R$ = 10. The most likely reason for this is that a large fraction of the radio emission arises in even more tenuous gas than the X-rays, mostly in the southwest. These fractions are very similar at 21~cm and 6~cm.

We consider that the radio emission correlated with the UV or X-rays must lie between the extreme $R$ = 1 and 10 (a visually reasonable solution is obtained for $R$ = 5), and we used them as an error interval. For each value of $R,$ we obtained the part of the radio emission that was correlated with the UV and that correlated with the X-rays. At 11 cm, the fraction of the radio associated with the UV is between 19\% ($R$ = 10) and 34\% ($R$ = 1), and that associated with the X-rays is between 3\% and 26\%. At all wavelengths the correlation with the UV is better than with the X-rays.

\section{Modeling the multiwavelength emission from the Cygnus Loop}
\label{section:Results}

To explain the observed $\gamma$-ray spectrum of the Cygnus Loop, we conducted a multiwavelength modeling of the remnant spectrum. Our analysis included radio data from 22~MHz up to 30~GHz (\citealt{Uyaniker_2004, Loru_2021} and citations therein), reduced by a constant factor reflecting the fraction of radio emission at 11 cm associated with the UV and/or X-ray emission, as explained in Section~\ref{section:radio_data}, together with the LAT GeV spectrum from this work. We modeled the radiative processes using the \texttt{naima} package~\citep{Zabalza_2015}. In our analysis, we assumed a distance to the SNR of $735$ pc~\citep{Fesen_2018}, and assuming a Sedov phase, a kinetic explosion energy of $E_{\rm SN} = 0.7 \times 10^{51}$ erg and an age of $t_{\rm age} \sim 21$ kyr.

We considered the contribution to the $\gamma$-ray spectrum from $\pi^0$ decay produced by the interactions of protons with ambient hydrogen, together with  contribution from Bremsstrahlung radiation and inverse Compton (IC) scattering by accelerated electrons, which also contribute to the radio through synchrotron emission. To take the presence of He, which has a spectral shape similar to protons in the spectrum of accelerated particles into account, we multiplied the emissivity from $\pi^0$ decay by a constant factor of 1.3. The ISM composition of the target gas is accounted for in \texttt{naima}.
Following~\citet{Katagiri_2011}, seed photons for IC scattering of electrons include the cosmic microwave background, two infrared ($T_{\rm IR}$ = 34, 470 K, $U_{\rm IR}$ = 0.34, 0.063 eV cm$^{-3}$, respectively), and two optical components ($T_{\rm opt} = 3.6 \times 10^3$, $9.9 \times 10^3$ K, $U_{\rm opt}$ = 0.45, 0.16 eV cm$^{-3}$, respectively) in the vicinity of the Cygnus Loop. Emission from the secondary electrons is neglected because of the low-density environment around the remnant.

\subsection{Ambient parameters}
\label{section:ambient}

\begin{table}
\caption{Model parameters for the Cygnus Loop.}
\label{tab:parameters}      
\begin{tabular}{ l r }     
\hline\hline 
Parameters & Values \\
\hline
\textbf{SNR dynamics}                                 &                     \\
Distance\tablefootmark{a}: $D$                        &   735 pc          \\
Radius\tablefootmark{a}: $R$                          &   18.5 pc         \\
Age\tablefootmark{a}: $t_{\rm age}$                     &   21000 yr        \\
Shock-cloud interaction\tablefootmark{b}: $t_{\rm c}$   &   $\sim$ 1200 yr    \\
Explosion Energy\tablefootmark{a}: $E_{\rm SN}$       & $7 \times 10^{50}$ erg \\
\hline
\multicolumn{2}{l}{\textbf{N-E region parameters (non-radiative shocks)}} \\
Density\tablefootmark{a}: $n_{\rm 0,cl}$                &   1.5 cm$^{-3}$    \\
Magnetic Field: $B_{\rm 0,cl}$                        &   3.0 $\mu$G        \\
Cloud shock velocity\tablefootmark{a}: $v_{\rm s}$     &   244 km s$^{-1}$  \\
Cutoff energy\tablefootmark{c}: $p_{\rm max}c$ ($\eta = 7$) & 15 GeV        \\
Break energy\tablefootmark{c}: $p_{\rm br}c$ ($T = 10^4$ K) &   62 GeV      \\
Energy in CRs\tablefootmark{c}: $W_{\rm tot}$      & $1.4 \times 10^{49}$ erg \\
e/p ratio\tablefootmark{c}: $K_{\rm ep}$ (at $10$ GeV)    &   0.01         \\
\hline
\multicolumn{2}{l}{\textbf{West region parameters (radiative shocks)}} \\
Density\tablefootmark{b}: $n_{\rm 0,cl}$               &   6 cm$^{-3}$      \\
Cooled gas density: $n_{\rm m}$                       &   293 cm$^{-3}$     \\
Magnetic Field\tablefootmark{b}: $B_{\rm 0,cl}$        &   6 $\mu$G         \\
Compressed Field: $B_{\rm m}$                         &   244 $\mu$G        \\
Cloud shock velocity\tablefootmark{b}: $v_{\rm s}$    &   130 km s$^{-1}$  \\
Cutoff energy\tablefootmark{c}: $p_{\rm max}c \times s^{1/3}$ ($\eta = 7$) & 20 GeV \\
Break energy\tablefootmark{c}: $p_{\rm br}c \times s^{1/3}$ ($T = 10^4$ K) & 70 GeV \\
Filling factor\tablefootmark{c}: $f$                 &   0.013           \\
\hline
\multicolumn{2}{l}{\textbf{Intercloud region}}                                         \\
Density\tablefootmark{a}: $n_{\rm 0}$                 &   0.4 cm$^{-3}$    \\
Magnetic Field: $B_{\rm 0}$                           &   2 $\mu$G         \\
Shock velocity\tablefootmark{a}: $v_{\rm s}$          &   350 km s$^{-1}$  \\

\hline
\end{tabular}
\tablefoot{}
\tablefoottext{a}{Taken from~\citealt{Fesen_2018}.}  \\
\tablefoottext{b}{Taken from~\citealt{Raymond_2020}.} \\
\tablefoottext{c}{Fit to the data.}
\end{table}
   
The Cygnus Loop blast wave has encountered discrete clouds to the east and northeast and a large molecular cloud to its west approximately $t_{\rm c} \sim 1200$ yr ago~\citep{Raymond_2020}. The range of shock speeds can vary widely in these regions due to the interaction of the remnant with the environment. In our analysis, we considered a cloud shock velocity $v_{\rm s} = 244$ km s$^{-1}$~\citep{Fesen_2018} and an upstream cloud density $n_{\rm 0,cl} = 1.5$ cm$^{-3}$~\citep{Long_1992}, where smooth nonradiative Balmer-dominated filaments are present. We assumed a cloud shock velocity $v_{\rm s} = 130$ km s$^{-1}$ and an upstream cloud density $n_{\rm 0,cl} = 6$ cm$^{-3}$~\citep{Raymond_2020}, where the deceleration is faster and UV and optical line emission cools down and compresses the gas, producing regions of radiative filaments. In between these dense clouds, the remnant expands in a low-density region~\citep[$\sim$ 0.4 cm$^{-3}$,][]{Raymond_2003} with a faster shock velocity~\citep[$\sim$ 350 km s$^{-1}$,][]{Medina_2014, Raymond_2015, Fesen_2018}.

To compute the required physical parameters in the cooled radiative regions of the remnant, we followed the approach described in~\citet{Uchiyama_2010}. The upstream magnetic field strength and density in the clouds are related by
\begin{equation}
\label{eq:B0cl}
    B_{\rm 0,cl} = b \sqrt{\frac{n_{\rm 0,cl}}{{\rm cm}^{-3}}} \; \mu G,\end{equation}
\noindent where $b = v_{\rm A}/(1.84$ km s$^{-1})$, with $v_{\rm A}$ being the Alfvén velocity, and ranges between $0.3$ and $3$~\citep{Hollenbach_1989}. \citet{Raymond_2020} found for the radiative regions an upstream magnetic field value of 6 $\mu$G , which following equation~\ref{eq:B0cl}, implies $b = 2.5$. Using the same $b$ value for the nonradiative shock regions, we found $B_{\rm 0,cl}$ = 3 $\mu$G. The magnetic field just downstream (before radiative compression, if any) is $B_{\rm d,cl} = r_{\rm B} \, B_{\rm 0,cl}$, where the magnetic compression ratio $r_{\rm B} = \sqrt{(2r_{\rm sh}^2+1)/3}$~\citep{Berezhko_2002} assumes a turbulent field ($r_{\rm sh} = 4$ is the shock compression ratio).

The density of the cooled gas in the radiative shocks, $n_{\rm m}$, was obtained by assuming that the compression is limited by magnetic pressure. Because the compression is strong, only the tangential field remains in the compressed magnetic field,
\begin{equation}
\label{eq:Bm}
B_{\rm m} = \sqrt{2/3}\, (n_{\rm m}/n_{\rm 0,cl}) \, B_{\rm 0,cl}    
.\end{equation}
We define $v_{\rm s7}$ the shock velocity in units of 100~km~s$^{-1}$. Equating $B_{\rm m}^2/8\pi$ with the shock ram pressure $n_{\rm 0,cl} \, \mu_{\rm H} \, v_{\rm s}^2$, where $\mu_{\rm H} \sim 1.4 \, m_{\rm p}$ is the mean mass per proton, we obtain
\begin{equation}
\label{eq:nm}
    n_{\rm m} \simeq 94 \, n_{\rm 0,cl} \, v_{\rm s7} \, b^{-1}
.\end{equation}
For the regions dominated by radiative shocks, we computed values of $n_{\rm m} = 293$ cm$^{-3}$ and $B_{\rm m} = 244$ $\mu$G \citep[consistent with those reported in][]{Raymond_2020}. A summary of the parameters we used can be found in Table~\ref{tab:parameters}.

\subsection{Particle spectrum}

\begin{figure*}
   \centering
            {\includegraphics[width=0.45\textwidth]{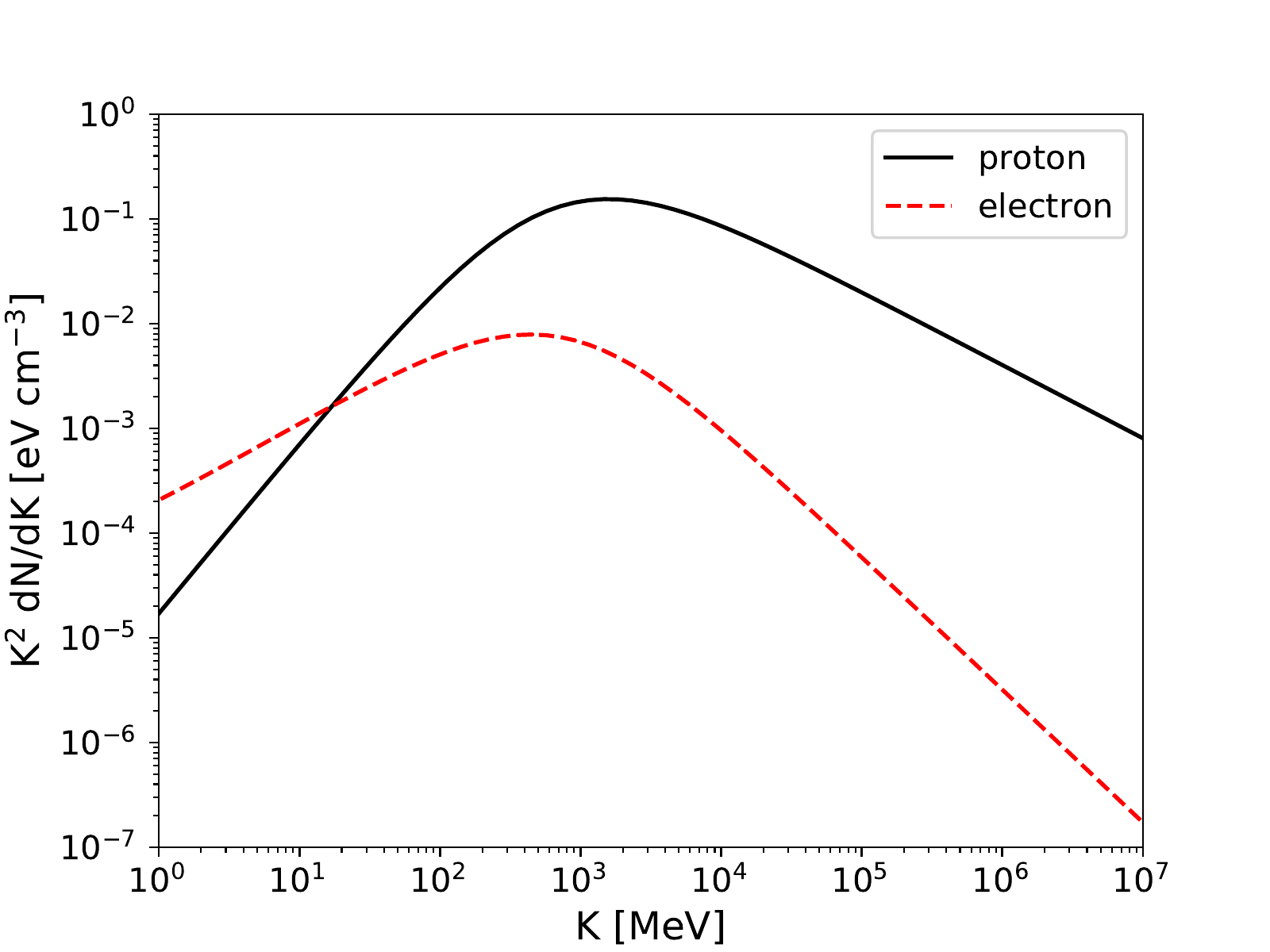}} \quad
            {\includegraphics[width=0.45\textwidth]{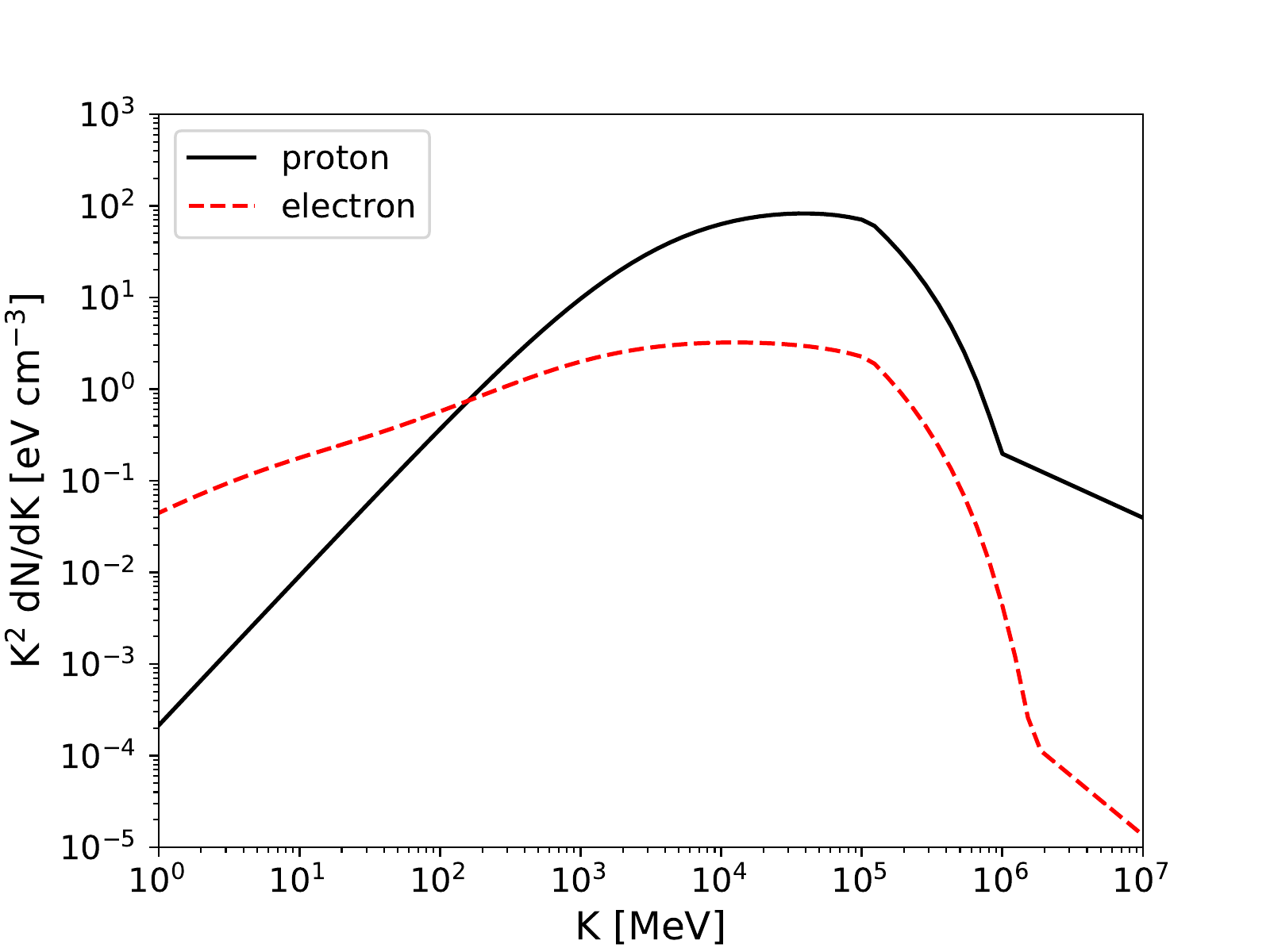}} \\
            
   \caption{Spectra of protons and electrons. \textit{Left:} Preexisting ambient CR spectra of protons (solid black) and electrons (dashed red) from~\citet{Phan_2018} as a function of the kinetic energy of the particles. \textit{Right:} Reaccelerated and compressed protons (solid black) and electrons (dashed red) vs kinetic energy with radiative compression $s = 12$, a cutoff at 100 GeV, and a break at 50 GeV as reference values.}
   \label{fig:CRs}
   \end{figure*}

We discuss the CR spectrum we used to model the multiwavelength emission from the Cygnus Loop. Two mechanisms can contribute to the observed emission: the diffusive shock acceleration (DSA) of thermal injected particles, and reacceleration of Galactic CRs (GCR).
We first discuss the model involving reacceleration of preexisting ambient CRs (hereafter RPCR). This model was adopted by~\citet{Uchiyama_2010} to explain $\gamma$-ray and radio emission from W~44. At the main shock location, the reaccelerated CRs number density $n_{\rm acc}(p)$ is given by
\begin{eqnarray}
  \label{eq:nacc}
n_{\rm int}(p) & = & (\delta + 2) \, p^{-\delta} \int_{0}^{p} n_{\rm GCR}(x) x^{\delta - 1} dx \\
n_{\rm acc}(p) & = &
\begin{cases} 
  \mathrm{e}^{-p/p_{\rm max}} \, n_{\rm int}(p), & p \leq p_{\rm br}, \\
  \frac{p_{\rm br}}{p} \, \mathrm{e}^{-p/p_{\rm max}} \, n_{\rm int}(p), & p \geq p_{\rm br}. 
\end{cases}
\end{eqnarray}
\noindent 
that is, the steady-state DSA spectrum $n_{\rm int}(p)$ \citep{Blandford_1987} with an exponential cutoff at $p_{\rm max}$ and a break at $p_{\rm br}$, where $\delta = \frac{r_{\rm sh}+2}{r_{\rm sh}-1}$ (in this work $\delta = 2$), $n_{\rm GCR}(p)$ is the preexisting ambient CR density and $p$ is the particle momentum.

We tried two parameterizations of the Galactic CR spectrum: We used the spectra of the Galactic CR protons and electrons from~\citet{Uchiyama_2010} and from~\citet{Phan_2018}, the former derived by data from \citet{Strong_2004} and \citet{Shikaze_2007}, the latter derived by fitting together local CR data from the Voyager 1 probe~\citep{Cummings_2016} and from the Alpha Magnetic Spectrometer \citep[AMS,][]{Aguilar_2015}.

In order to take the maximum attainable energy of particles (due to energy loss and finite acceleration time) into account,we introduced an exponential cutoff at $p_{\rm max}$. Following~\citet{Uchiyama_2010}, the age-limited maximum momentum is
\begin{equation}
\label{eq:pmax}
p_{\rm max} = 50 \, (\eta/10)^{-1} \, v_{\rm s7}^2 \, t_4 \, (B_{\rm 0,cl}/10 \mu G)  \; {\rm GeV/c}
,\end{equation}
\noindent where $t_{\rm 4}$ is the remnant age (or the shock-cloud interaction age $t_{\rm c}$) in units of $10^4$ yr. The gyro or Bohm factor $\eta$ depends on the remnant age; it is $\eta \sim 1$ for efficient and young SNRs such as RX J1713.7$-$3946~\citep{Uchiyama_2007, Tsuji_2019}, but larger than 1 in older SNRs ($\eta = 10$ in \citealt{Uchiyama_2010}).

We also considered a spectral steepening above $p_{\rm br}$ for both electrons and protons. The cooling break in the electron population was calculated by equating the synchrotron loss time~\citep{Parizot_2006},
\begin{equation}
\label{eq:sync}
    \tau_{\rm sync} = 1.25 \times 10^3 \, E_{\rm TeV}^{-1} \, (B_{\rm m}/100 \, \mu G)^{-2} \; {\rm yr}
,\end{equation}
\noindent and the remnant (or shock-cloud interaction) age. For the protons, we considered a spectral break at $p_{\rm br}$ induced by neutral-ion collisions~\citep{Malkov_2011},
\begin{equation}
    \label{eq:pbr}
    p_{\rm br} = \frac{2 \, e \, B_{\rm 0,cl} \, v_{\rm A}}{c \, \nu_{\rm i-n}}
,\end{equation}
\noindent
where $\nu_{\rm i-n} \simeq 9 \times 10^{-9} n_{\rm 0,cl} \, T_{\rm 4}^{\rm 0.4}$ s$^{-1}$ is the ion-neutral collision frequency and $T_{\rm 4}$ is the precursor temperature in units of $10^4$K.

Because of the adiabatic compression in the radiative shocks, each particle gains energy as $p \rightarrow s^{1/3}p$, where $s \equiv (n_{\rm m}/n_{\rm 0,cl})/r_{\rm sh}$ ($s = 12.22$ in this work).
Therefore the number density of accelerated and compressed CRs at the point where the density becomes $\sim n_{\rm m}$ is~\citep{Uchiyama_2010}
\begin{equation}
\label{eq:ad}
n_{\rm ad}(p) = s^{2/3}n_{\rm acc}(s^{-1/3}p)
.\end{equation}

The effect of reacceleration and compression on the CR proton and electron spectra based on eq.~\ref{eq:ad} is shown in Figure~\ref{fig:CRs}.
Following \citet{Uchiyama_2010}, we parameterized the emission volume as $V = f \, 4/3 \pi R^3$, where $f$ is the filling factor of the clouds before they were crushed with respect to the entire SNR volume. The particle spectrum integrated over the SNR volume is therefore
\begin{equation}
\label{eq:reacc}
N(p) = \left(\frac{n_{\rm 0}}{n_{\rm m}}\right) \, f \, \frac{4}{3}\pi R^3 \, n_{\rm ad}(p)
.\end{equation}
   
We then also considered the contribution of freshly accelerated CRs at the blast wave, according to the DSA theory.
The CR spectrum resulting from DSA of thermal injected particles for both protons and electrons is assumed to be a steady-state DSA spectrum with a break and an exponential cutoff given by eqs.~\ref{eq:pbr} and \ref{eq:pmax}.

\subsection{Nonradiative regions: DSA scenario}
\label{section:ne}

   \begin{figure}

            \includegraphics[width=\hsize]{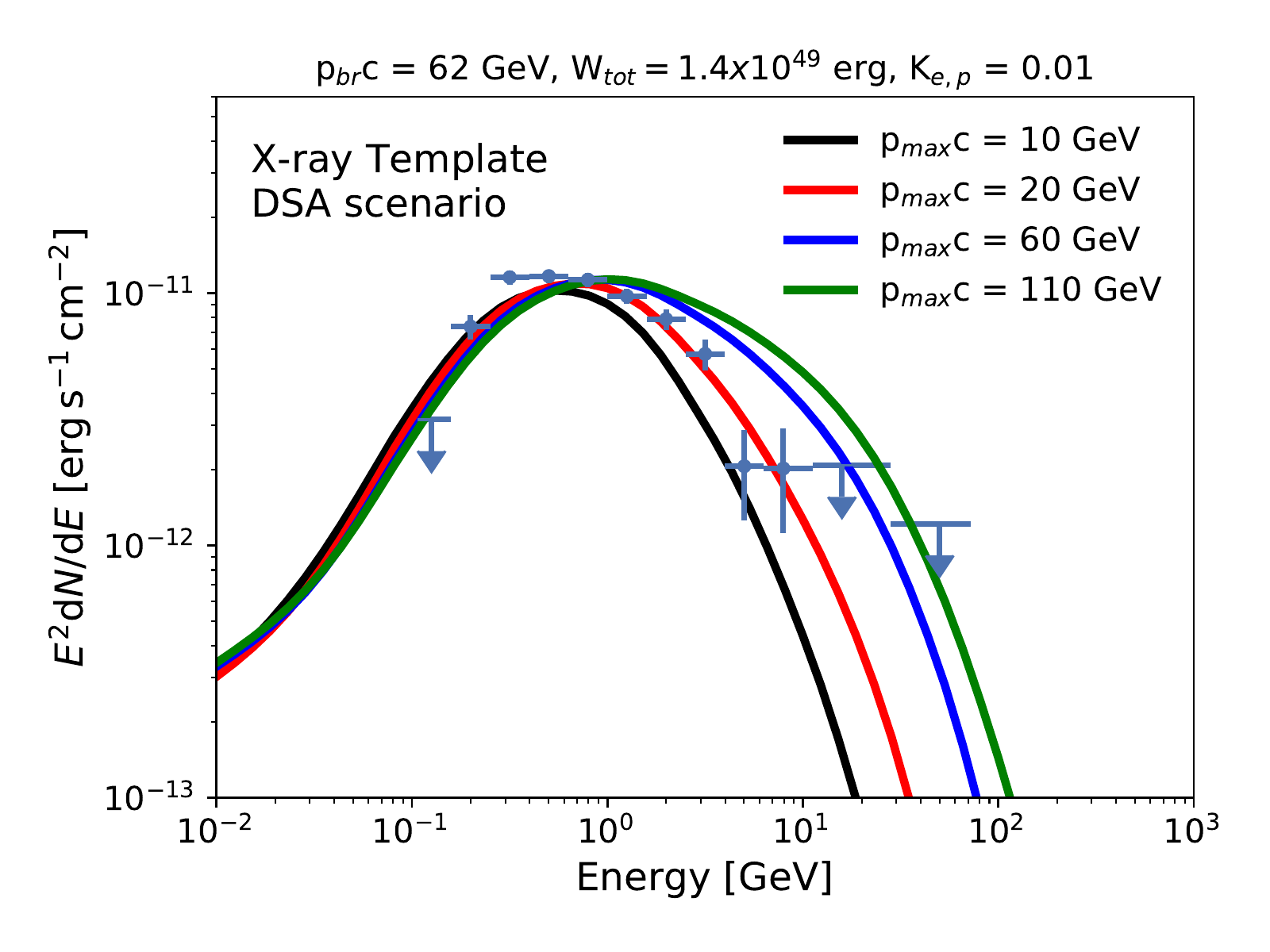}
            
   \caption{$\gamma$-ray spectrum of the Cygnus Loop extracted with the X-ray template and modeled emission in the DSA scenario. Solid lines show the total contribution of Bremsstrahlung radiation, IC, and $\pi^0$-decay. The total energy ($W_{\rm tot} = 1.2 \times 10^{49}$ erg), the electron-to-proton ratio ($K_{\rm ep} = 0.01$ at $10$ GeV), and the break energy ($p_{\rm br}$ = 62 GeV/c) are fixed. Different cutoff energy values are shown.}
   \label{fig:dsa_ne}
   \end{figure}

   \begin{figure}
   \centering
   \includegraphics[width=\hsize]{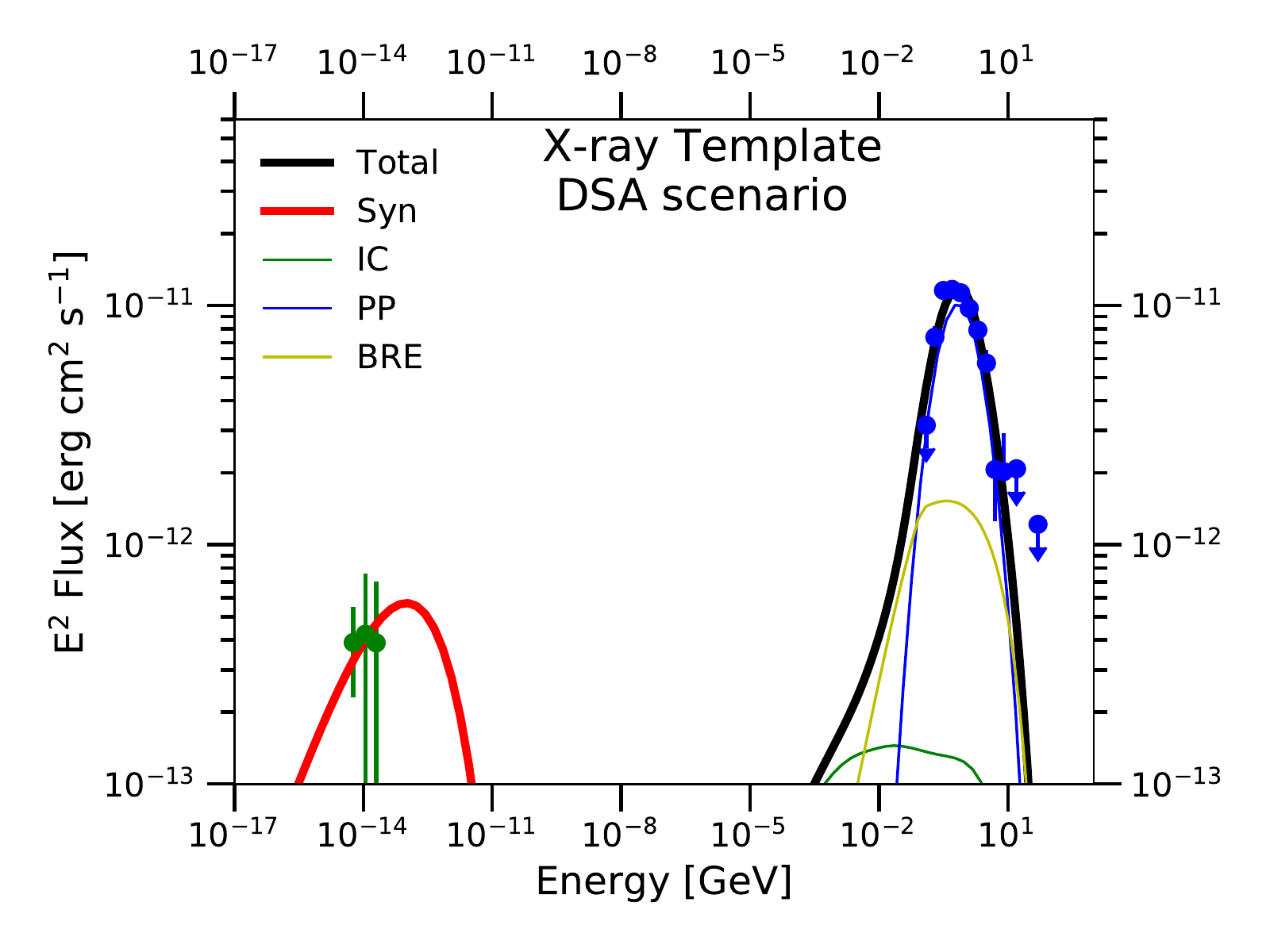}
      \caption{DSA scenario for the multiwavelength modeling of the emission, extracted with the X-ray template, toward the Cygnus Loop. Radio data are obtained as explained in Section \ref{section:radio_data}. The total contribution from Bremsstrahlung, IC, and $\pi^0$ decay is shown by the solid black line. Our best model is obtained with a total energy $W_{\rm tot}$ of $1.2 \times 10^{49}$ erg and the electron-to-proton ratio $K_{\rm ep} = 0.01$ at $10$ GeV. Protons and electrons have an energy cutoff of $15$ GeV and an energy break of $62$ GeV.}
         \label{fig:dsa_ne_best}
   \end{figure}
   
The nonradiative shocks of the Cygnus Loop are characterized by shocks that are fast enough to accelerate particles through the DSA mechanism. As we showed in Figures~\ref{fig:contour} and~\ref{fig:excess}, we described the $\gamma$-ray emission with a two-component model (X-ray+UV templates) in which the X-ray emission arises from the fast nonradiative shocks. We therefore modeled the emission extracted with the X-ray component (radio data extraction is described in Section~\ref{section:radio_data}, $\gamma$ -ray data are shown in lower panel of Figure~\ref{fig:seds}) using a particle distribution arising from the DSA mechanism. On the one hand, the environmental parameters are kept fixed in our model; on the other hand, the spectral parameters (the cutoff and break energies) depend on unknown parameters such as $\eta$ and $T_{\rm 4}$.

The environmental parameters that are best suited to model the X-ray related emission should be those of the intercloud region ($v_{\rm s}$ = 350 km s$^{-1}$, $n_{\rm 0}$ = 0.4 cm$^{-3}$), where the shock is fast enough to generate X-ray emission. However, when a pre-shock magnetic field of $B_{\rm 0}$ = 2 $\mu$G (see eq.~\ref{eq:B0cl}) is considered, eq.~\ref{eq:pmax} yields $p_{\rm max}c >$ 260 GeV ($t_{\rm 4} = t_{\rm age}, \eta = 10$), which is incompatible with the soft $\gamma$-ray emission. We therefore decided to use intermediate values ($v_{\rm s}$ = 244 km s$^{-1}$, $n_{\rm 0,cl}$ = 1.5 cm$^{-3}$, $B_{0,cl}$ = 3.0 $\mu$G, $t=t_{\rm c}$) to better fit the data points. Following equations~\ref{eq:pmax} and~\ref{eq:pbr}, the cutoff energy is $10 < p_{\rm max}c < 105$ GeV for $10 > \eta > 1$, and the energy break is $33 < p_{\rm br}c < 62$ GeV for $10^5 > T > 10^4$ K. Here, the break in the electron population can be neglected because synchrotron cooling is not relevant (see eq.~\ref{eq:sync}).

In Figure~\ref{fig:dsa_ne} we present the $\gamma$-ray spectrum from the Cygnus Loop, and we demonstrate the expected level of the $\gamma$-ray emission with varying $p_{\rm max}c$. To compute the $\gamma$-ray emission from $\pi^0$-decay, we used as the target density the upstream cloud density (1.5 cm$^{-3}$), as an average over the entire volume where cosmic rays are present. We kept fixed the total energy ($W_{\rm tot} = W_{\rm p} + W_{\rm He}$, where $W_{\rm p}$ and $W_{\rm He}$ are the total energy of protons and He, respectively) to $W_{\rm tot} = 1.2 \times 10^{49}$ erg (corresponding to $\sim 2\%$ of $E_{\rm SN}$) and the electron-to-proton differential spectrum ratio in kinetic energy $K_{\rm ep} = 0.01$ at $10$ GeV. Figure~\ref{fig:dsa_ne} shows the effect of the cutoff energy on the modeled emission. Because a low value of $p_{\rm max}c$ is necessary to fit the data, $p_{\rm br}c$ does not affect the model. Hence, we decided to set $p_{\rm max}c = 15$ GeV and $p_{\rm br}c = 62$ GeV (see Table~\ref{tab:parameters}), which correspond to $\eta = 7$ and $T = 10^4$ K, in order to reproduce the $\gamma$-ray data as shown in Figure~\ref{fig:dsa_ne_best}.
   
\subsection{Radiative regions: Reacceleration of preexisting ambient CRs}
\label{section:west}

   \begin{figure*}
   \centering
            {\includegraphics[width=0.45\textwidth]{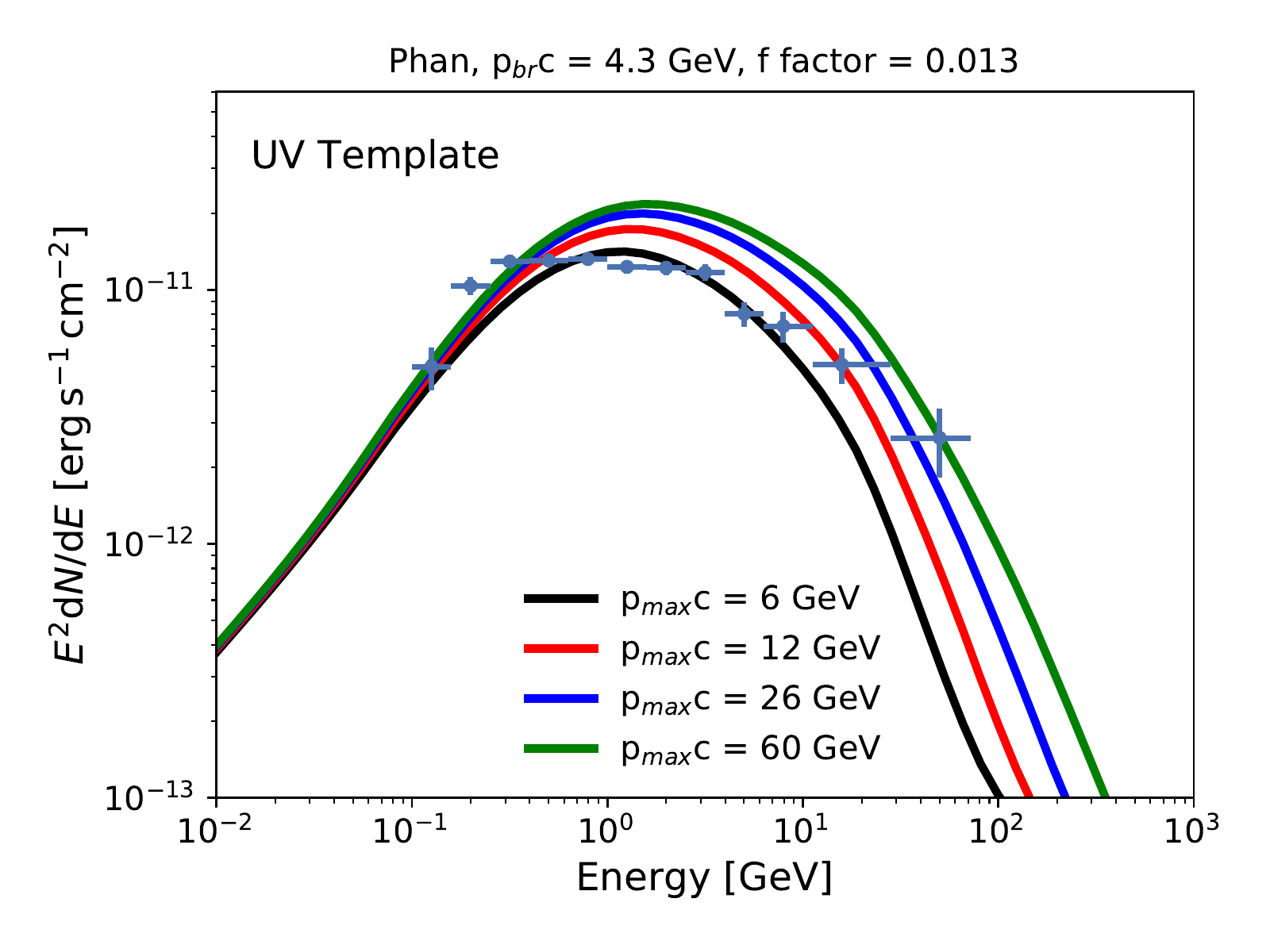}} \quad
            {\includegraphics[width=0.45\textwidth]{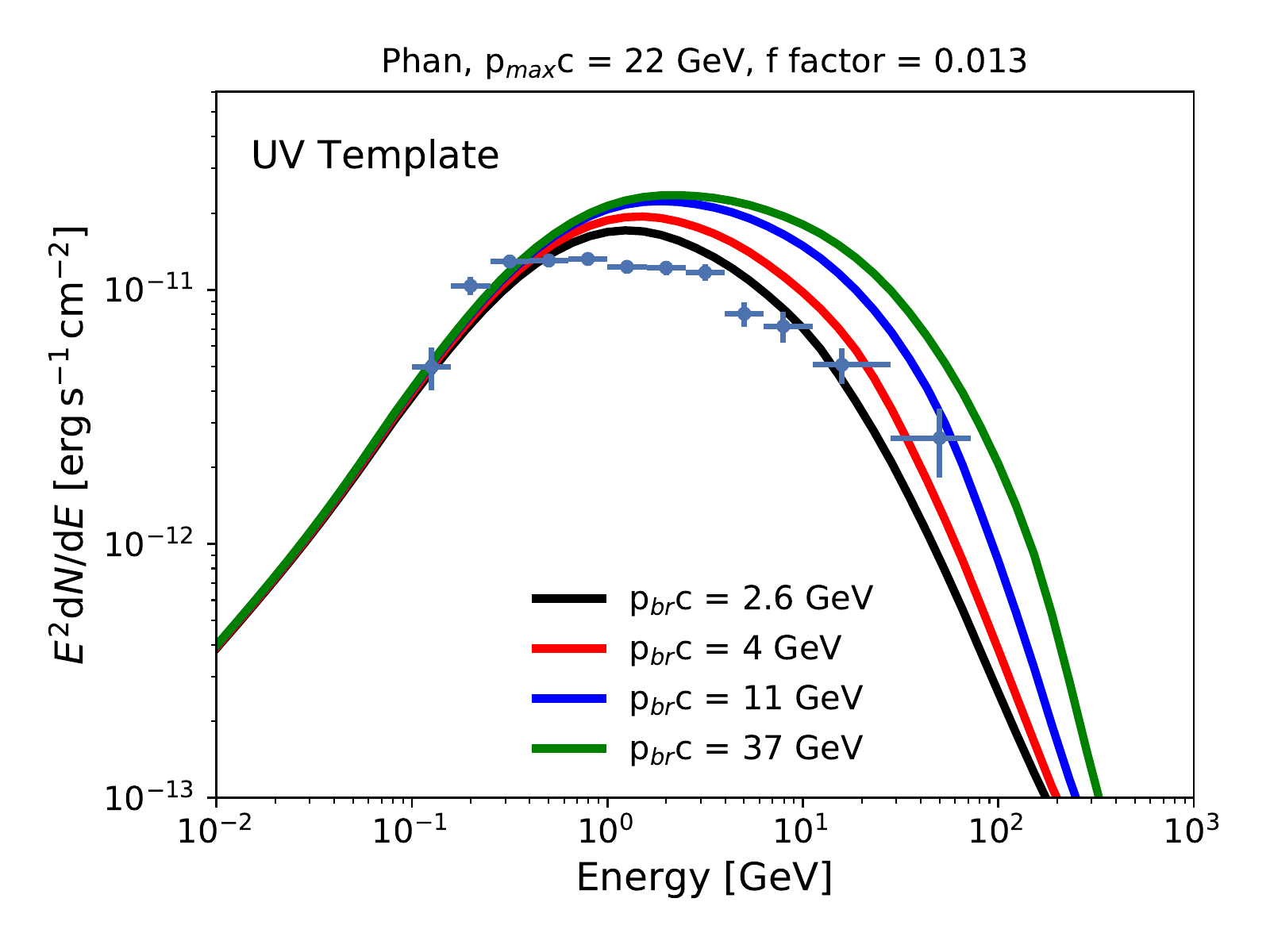}} \\
   \caption{$\gamma$-ray spectrum of the Cygnus Loop extracted with the UV template and modeled emission in the RPCR scenario (eq.~\ref{eq:reacc}). Solid lines show the total contribution of Bremsstrahlung radiation, IC, and $\pi^0$ decay using preexisting CRs from \citet{Phan_2018}. On the left (right), we show different cutoff (break) energy values at $p_{\rm br}c = 4.3$ GeV ($p_{\rm max}c = 22$ GeV).  The actual cutoff and break energies in the radiative zone are $s^{1/3} p_{\rm max}c$ and $s^{1/3} p_{\rm br}c$. The filling factor $f$ is fixed to $f = 0.013$.}
   \label{fig:reacc_west}
   \end{figure*}
   
   \begin{figure}
   \centering
   \includegraphics[width=\hsize]{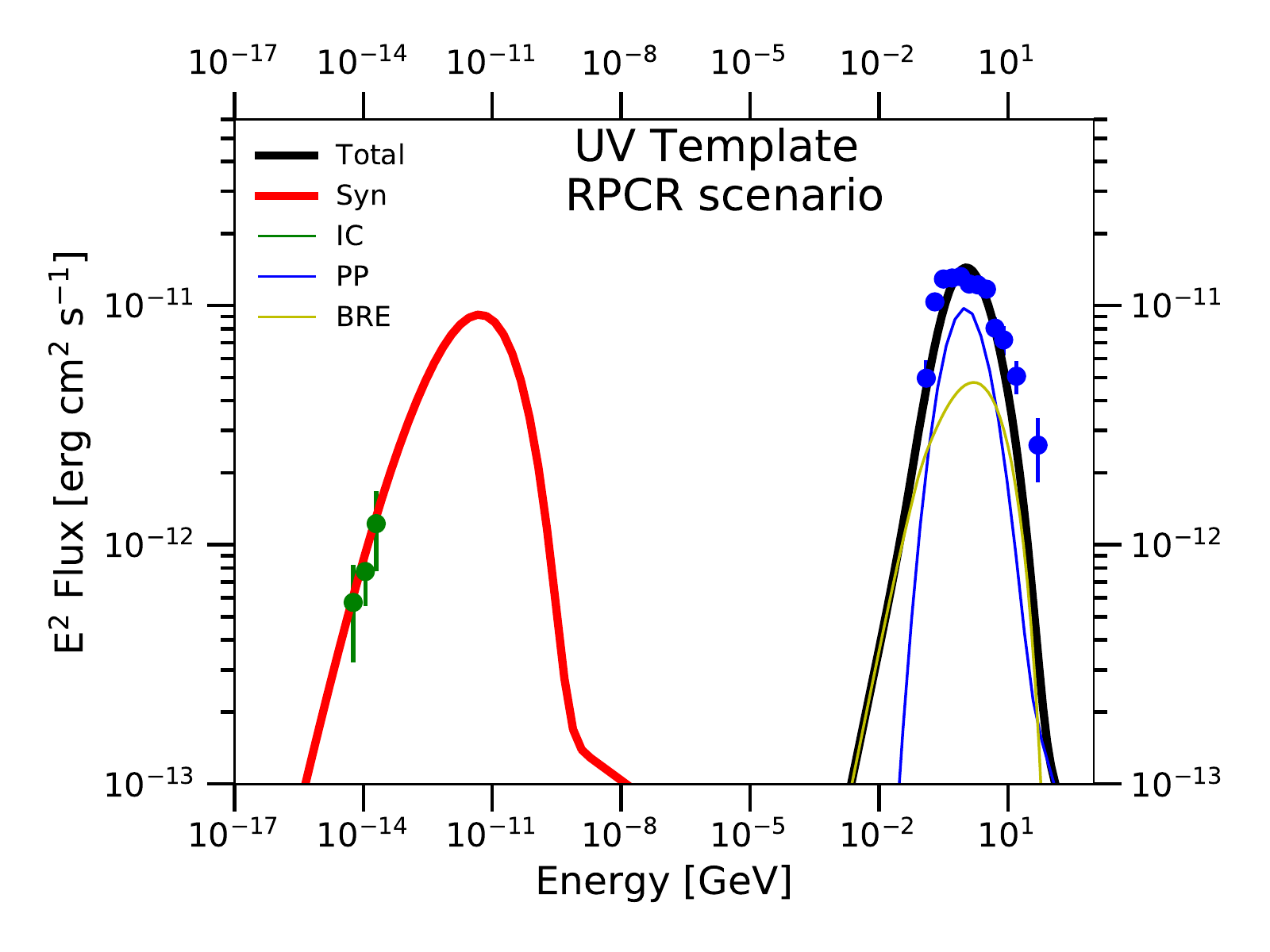}
      \caption{RPCR scenario for the multiwavelength modeling of the emission, extracted with the UV template, toward the Cygnus Loop. Radio data are obtained as explained in Section \ref{section:radio_data}. The total contribution from Bremsstrahlung, IC, and $\pi^0$ decay is shown by the solid black line. Our best model is obtained with preexisting CR populations from \citet{Phan_2018} and a filling factor of $f = 0.013$. Spectral parameters for protons and electrons are $s^{1/3} p_{\rm max}c = 20$ GeV and $s^{1/3} p_{\rm br}c = 70$ GeV.}
         \label{fig:reacc_west_best}
   \end{figure}

   In contrast to the nonradiative shocks, the radiative shocks are slower and cannot efficiently accelerate particles through the DSA mechanism. We then considered a model involving the RPCR  in regions dominated by radiative shocks. In Section~\ref{section:Analysis} we showed that part of the $\gamma$-ray emission of the remnant is associated with the UV component (emitted by radiative shocks) in the X-ray+UV model. We then used the SED data points extracted with the UV component (radio data extraction is described in Section~\ref{section:radio_data}, $\gamma$ - ray data are shown in lower panel of Figure~\ref{fig:seds}) to model these regions.

   Again, the spectral parameters of the compressed and reaccelerated particle populations are not constrained; we therefore explored values of $14 < s^{1/3} p_{\rm max}c < 140$ GeV for $10 > \eta > 1$ and $6 < s^{1/3} p_{\rm br}c  < 70$ GeV for $10^5 > T > 10^4$ K. The break due to synchrotron losses can be neglected. Another free parameter is the filling factor $f$ of the clouds, and it is obtained from the data.

   We explored two different preexisting ambient CR spectra: the Galactic CR protons, and electrons from \citet{Uchiyama_2010} and from \citet{Phan_2018}.
   By exploring different values of $p_{\rm max}$ and $p_{\rm br}$ for both preexisting CR spectra, we found that the differences between the two reaccelerated particle populations are minimal,  also in terms of $\gamma$-ray emission. We therefore decided to use preexisting CRs from \citet{Phan_2018}, obtained from the more recent Voyager 1 \citep{Cummings_2016} and AMS-02 \citep{Aguilar_2015} data. In Figure~\ref{fig:reacc_west}, as in Figure~\ref{fig:dsa_ne}, we present the expected level of the $\gamma$-ray emission with varying $p_{\rm max}c$ and $p_{\rm br}$. To describe the data, we used $s^{1/3} p_{\rm max}c = 20$ GeV (corresponding to $\eta = 7$) and $s^{1/3} p_{\rm br}c = 70$ GeV (corresponding to $T = 10^4$ K). The best filling factor is $f = 0.013$ (see Table~\ref{tab:parameters}). The resulting fit to the spectrum of the radiative regions is shown in Figure~\ref{fig:reacc_west_best}. Our model is too peaked in $\gamma$ rays and fails to fit the data at energies > 10 GeV.
   
\subsection{Modeling the entire Cygnus Loop}

   \begin{figure}
   \centering
            {\includegraphics[width=0.45\textwidth]{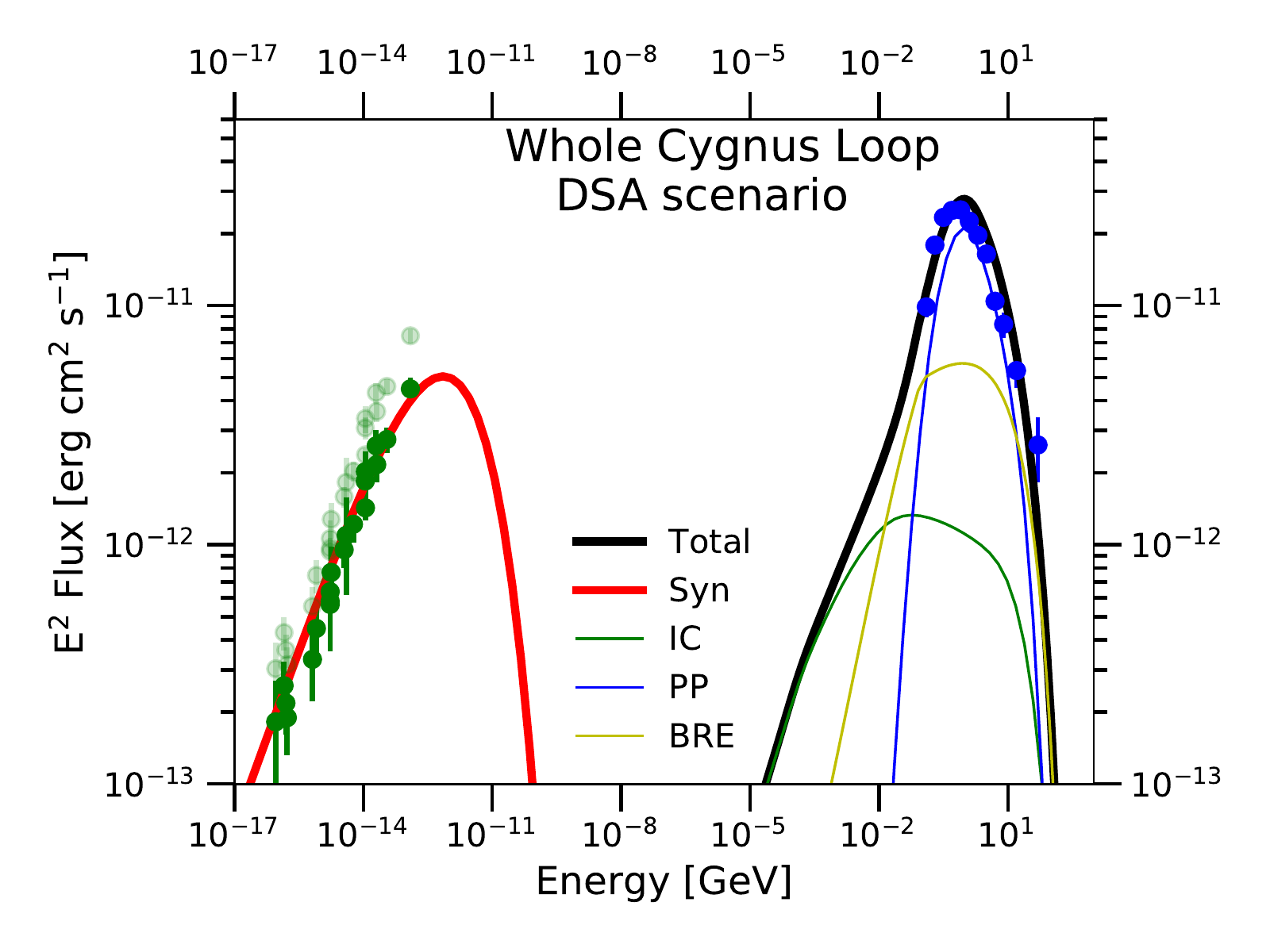}} \quad
            {\includegraphics[width=0.45\textwidth]{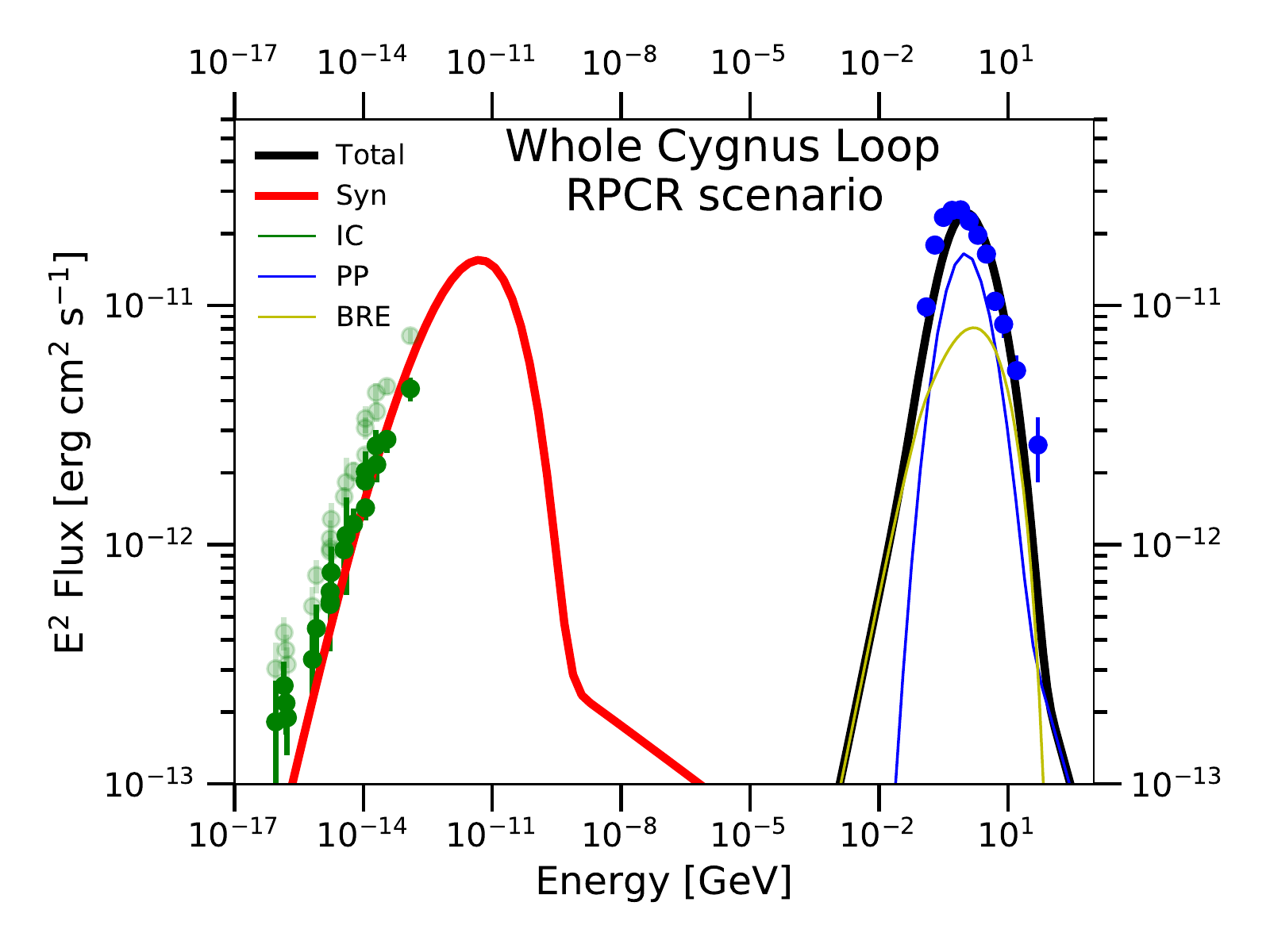}} \quad
            
   \caption{Radio and $\gamma$-ray emission from the entire Cygnus Loop. Radio data points (transparent green) from \citet{Uyaniker_2004} and \citet{Loru_2021} were scaled by a constant factor (dark green) to remove the contribution from the southern region (see Section \ref{section:radio_data}). The total contribution from Bremsstrahlung, IC, and $\pi^0$ decay is shown by the solid black line. \textit{Upper panel:} DSA scenario, adopting values of $p_{\rm max}c = 40$ GeV, $p_{\rm br}c = 62$ GeV, $W_{\rm tot} = 2.1 \times 10^{49}$ erg, and $K_{\rm ep} = 0.025$ for protons and electrons. \textit{Lower panel:} RPCR scenario, adopting preexisting CR populations from \citet{Phan_2018}, $s^{1/3} p_{\rm max}c = 20$ GeV, $s^{1/3} p_{\rm br}c = 70$ GeV, and a filling factor $f \sim 0.02$.}
   \label{fig:all_test}
   \end{figure}
   
   \begin{figure}
   \centering
            {\includegraphics[width=0.45\textwidth]{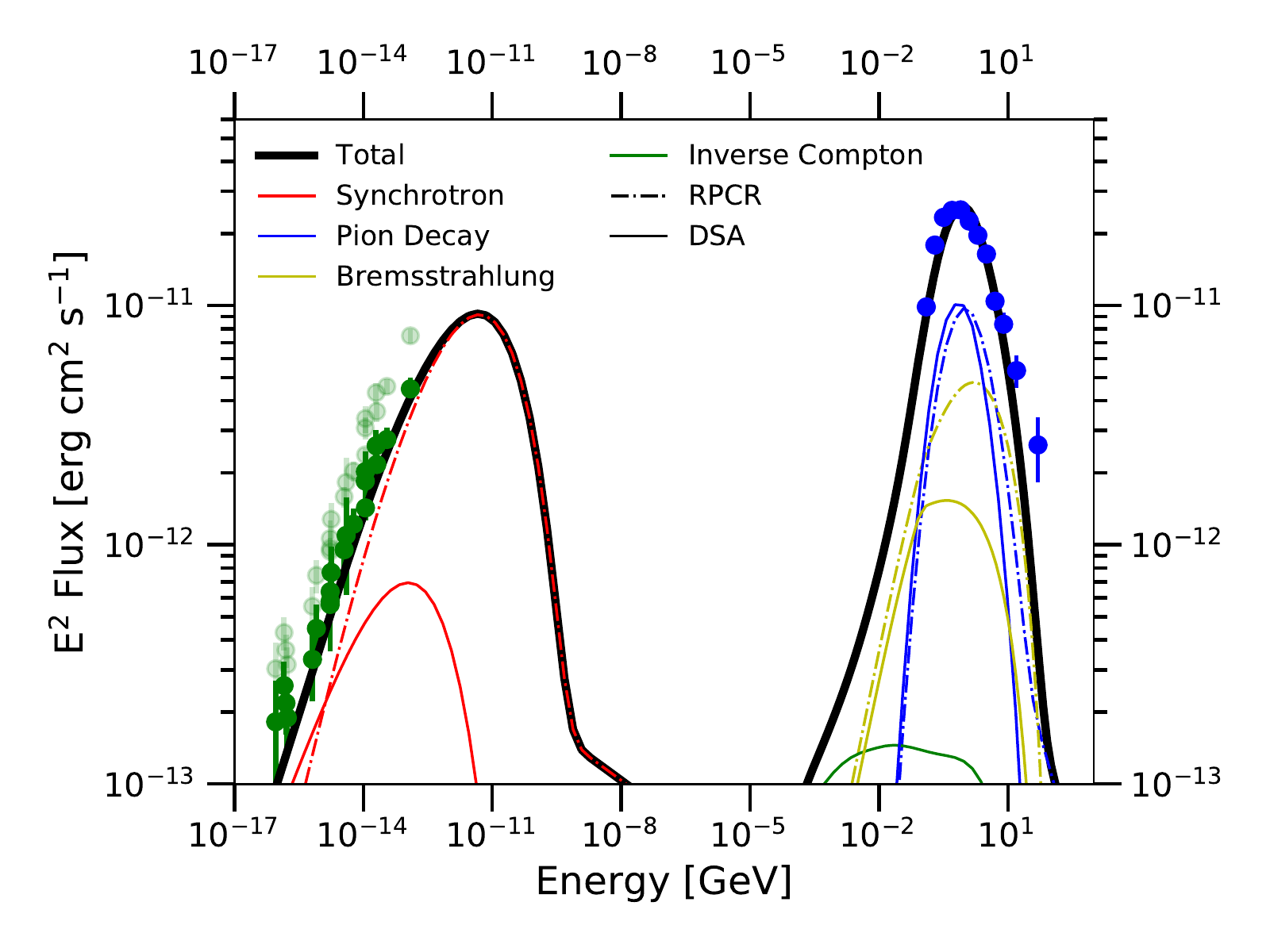}} \quad
            {\includegraphics[width=0.45\textwidth]{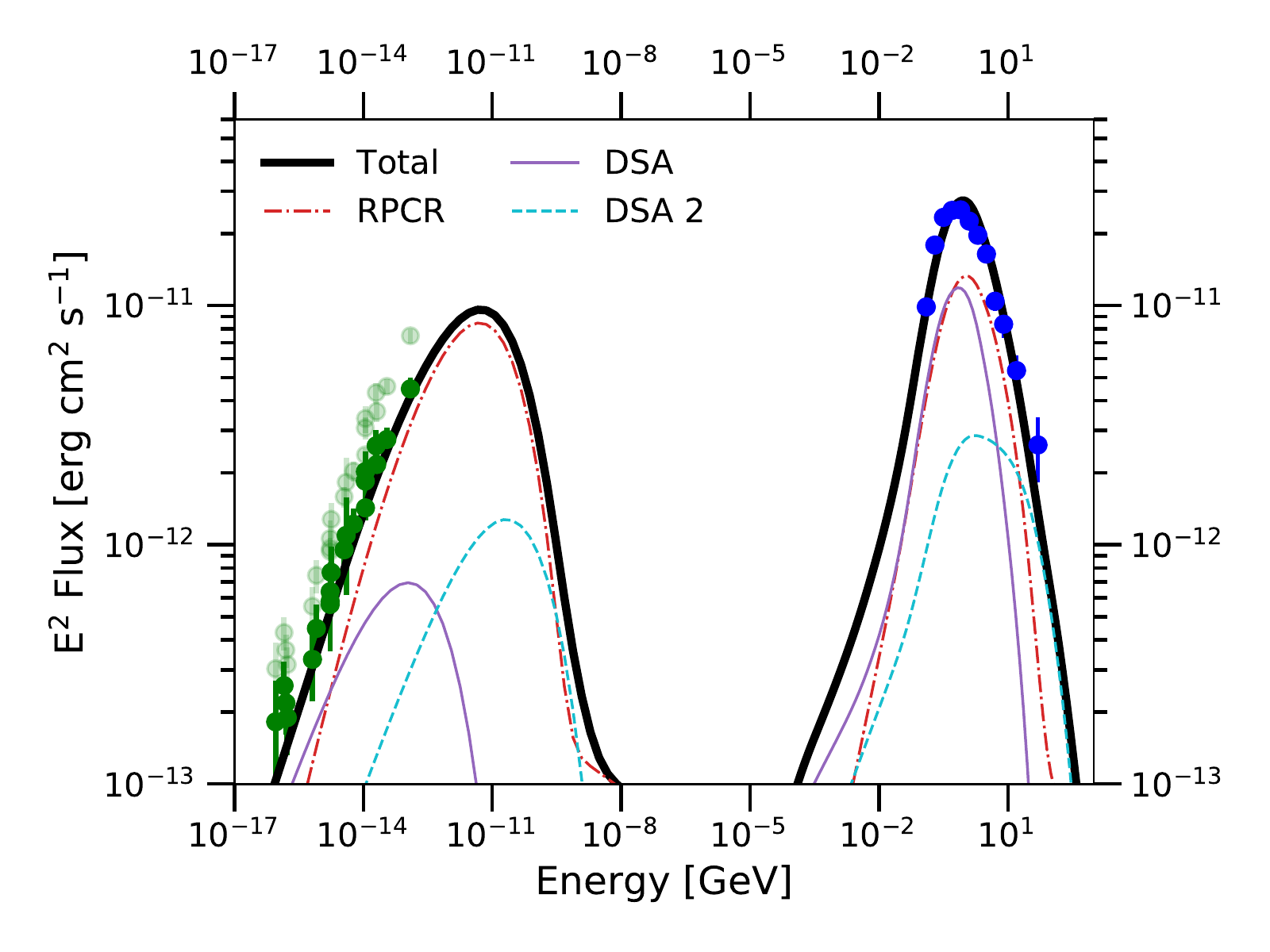}} \quad
      \caption{Same as Figure~\ref{fig:all_test}, but the multiwavelength model has multiple contributions. \textit{Upper panel:} DSA scenario (solid lines) and RPCR scenario (dot-dashed lines). The total contribution from the two scenarios is shown by the solid black line. \textit{Lower panel:} Same as the upper panel, but with contribution from the intercloud region (DSA 2).}
         \label{fig:all_best}
   \end{figure}

   We first attempted to model the emission from the entire Cygnus Loop (obtained using the UV template alone), considering either a DSA or an RPCR scenario. Assuming the same environmental parameters as for the northeast region ($v_{\rm s} = 244$ km $s^{-1}$, $n_{\rm 0,cl} = 1.5$, $B_{\rm 0,cl} = 3.0 \mu$G), we tried to model the multiwavelength emission in a DSA scenario. When values of $p_{\rm max}c = 40$ GeV and $p_{\rm br}c = 62$ GeV for protons and electrons together with $W_{\rm tot} = 2.1 \times 10^{49}$ erg and $K_{\rm ep} = 0.025$ are adopted, the $\gamma$-ray spectrum can be well reproduced by the DSA model shown in Figure~\ref{fig:all_test} (upper panel). This energy cutoff requires $\eta = 3$ (see eq.~\ref{eq:pmax}), which is lower than the typical $\eta = 10$ found in other intermediate-age SNRs~\citep{Uchiyama_2010, Devin_2018, Devin_2020, Abdollahi_2020b}.
   
In addition, it clearly emerged from our morphological analysis (see Section~\ref{section:Analysis}) that the $\gamma$-ray emission is mainly correlated with the UV templates and a consequence, with the radiative regions. Therefore the DSA mechanism is not favored to explain the Cygnus Loop multiwavelength emission.

We also tried to model the overall spectrum assuming an RPCR scenario. We found that the best parameterization to model the overall spectrum is the same as that reported in Section~\ref{section:west} ($s^{1/3} p_{\rm max}c = 20$ GeV, $s^{1/3} p_{\rm br}c = 70$ GeV, and preexisting CRs from~\citealt{Phan_2018}), except for $f \sim 0.02$. However, this model presents several discrepancies with observed data, as shown in the lower panel of Figure~\ref{fig:all_test}. The synchrotron emission is not able to reproduce the radio points at lower energies,
and the peak in the $\gamma$-ray emission (at $\sim$ 2 GeV) is higher than in the LAT data. Therefore the RPCR scenario alone is not able to satisfactorily explain the emission from the entire remnant either.

From our analysis reported in Sections~\ref{section:ne} and~\ref{section:west}, it emerged clearly that the nonthermal emission from the radiative and nonradiative regions has a different physical origin. As a consequence, we propose to model the total SNR spectrum with two contributions: an RPCR scenario caused by radiative shocks arising from the denser clouds, and a DSA contribution connected to the faster shock traveling in the lower density environment.

The upper panel of Figure~\ref{fig:all_best} shows the corresponding emission model with the contribution from the DSA (solid lines) and the RPCR (dot-dashed lines). The parameters are set exactly like those reported in Sections~\ref{section:ne} and~\ref{section:west} for the DSA and RPCR contribution, respectively (see Table~\ref{tab:parameters}). The high magnetic field in the cooled regions behind the radiative shocks makes the contribution of the RPCR scenario dominant with respect to the DSA in the radio band, while in the $\gamma$-ray band, the two components have similar contributions, reflecting the $\gamma$-ray flux associated with the X-ray and UV templates (see Section~\ref{section:spectral}). Overall, the modeled emission reproduces the observed data points in the radio and $\gamma$-ray bands well, unveiling the complex origin of the nonthermal emission of the remnant.

The model is slightly too soft to fit the highest $\gamma$-ray energy points. A contribution at these energies could come from particles that are accelerated, through DSA, by the faster shocks in the low-density intercloud medium. This scenario (namely, DSA 2) could arise from the parameters described previously ($v_{\rm s} = 350$ km $s^{-1}$, $n_{\rm 0} = 0.4$, and $B_{\rm 0} = 2 \mu$G), implying $p_{\rm max}c \sim 260$ GeV. We set the total energy of the protons and the electron-proton ratio equal to those of the DSA component (i.e., $W_{\rm tot} = 1.2 \times 10^{49}$ erg and $K_{\rm ep} = 0.01$). This new component is shown in the lower panel of Figure~\ref{fig:all_best}. By adding this new component, our model is able to explain the entire spectrum. When compared to the previous model, adding the DSA 2 component gives a TS value of 12 that is computed from the SED points.

\section{Conclusions}
\label{section:Conclusions}

We have presented the analysis of $\sim 11$ years of \textit{Fermi}-LAT data in the region of the Cygnus Loop. Our morphological analysis between 0.1 and 100 GeV confirmed an extended emission in the $\gamma$-ray band in the shape of a ring with maximum and minimum radii of 1.50$\degr$ (+0.01,$-$0.02) and 0.50$\degr$ (+0.04,$-$0.07), respectively. We found a strong correlation between the $\gamma$-ray emission and the X-ray and UV thermal emission. In particular, we found that the GeV morphology of the Cygnus Loop is best described by a two-component model: one consisting of a spatial template obtained from the X-ray thermal emission that is brightest in the northeast region of the remnant, the other consisting of a UV spatial template that dominates the central and west regions of the remnant. The $\gamma$-ray spectra extracted from these two components present a peak at $\sim 1$ GeV and can be described by the LogParabola function. Overall, the Cygnus Loop has a $\gamma$-ray spectrum that can be described by a power law with subexponential cutoff toward low energies and an integrated energy flux in the energy band  0.1 -- 100 GeV of $9.0 \pm 0.2 \times 10^{-11}$ erg cm$^{-2}$ s$^{-1}$.

The peak in the $\gamma$-ray spectrum suggests a hadronic origin of the nonthermal emission, as already shown by~\citet{Katagiri_2011}. We constrained the high-energy particle population using the radio and $\gamma$-ray emission. The wide range of shock speeds in different regions in the Cygnus Loop together with the results of our morphological analysis indicates two different possible physical scenarios for the origin of these particles: the DSA mechanism in regions with shock velocity $> 150$ km s$^{-1}$, and RPCR otherwise. 
Our multiwavelength analysis confirms that neither scenario alone is capable of explaining the entire nonthermal emission from the Cygnus Loop, but a model involving both scenarios simultaneously works well. We found that two different populations of hadrons and leptons are responsible for the nonthermal emission: one arising from the DSA mechanism, the other due to the RPCR.

Our best-fit model requires a maximum attainable energy of $\sim 15$ GeV for hadrons and leptons in the DSA and RPCR populations. In this model,  $2\%$ of the kinetic energy released by the SN go into particles accelerated through DSA (another fraction could have already escaped), and an electron-proton ratio of K$_{\rm ep} \sim 0.01$. The pre-shock filling factor for the RPCR scenario is $< 0.02$. Because the particles have a harder spectrum below the cutoff in the RPCR scenario, the RPCR component in the $\gamma$-ray spectrum is harder than the DSA component. By extracting radio and $\gamma$-ray contributions from the entire Cygnus Loop using the X-ray and UV templates, we disentangled the two different contributions to the nonthermal emission and unveiled the multiple origins of the accelerated particles in the remnant.

Although it has been studied for many years, the Cygnus Loop continues to be of great interest to the community. 
Models describing the full evolution of the remnant~\citep{Ferrand_2019, Ono_2020, Orlando_2020, Tutone_2020} and its thermal and nonthermal emission~\citep{Orlando_2012, Miceli_2016, Orlando_2019, Ustamujic_2021} would be very useful.

\begin{acknowledgements}
The \textit{Fermi} LAT Collaboration acknowledges generous ongoing support from a number of agencies and institutes that have supported both the development and the operation of the LAT as well as scientific data analysis. These include the National Aeronautics and Space Administration and the Department of Energy in the United States, the Commissariat à l’Energie Atomique and the Centre National de la Recherche Scientifique / Institut National de Physique Nucléaire et de Physique des Particules in France, the Agenzia Spaziale Italiana and the Istituto Nazionale di Fisica Nucleare in Italy, the Ministry of Education, Culture, Sports, Science and Technology (MEXT), High Energy Accelerator Research Organization (KEK) and Japan Aerospace Exploration Agency (JAXA) in Japan, and the K. A. Wallenberg Foundation, the Swedish Research Council and the Swedish National Space Board in Sweden. Additional support for science analysis during the operations phase is gratefully acknowledged from the Istituto Nazionale di Astrofisica in Italy and the Centre National d’Études Spatiales in France. This work performed in part under DOE Contract DE- AC02-76SF00515.
\end{acknowledgements}

\bibliography{Biblio}
\bibliographystyle{aa}

\end{document}